\documentclass{aa} 

\usepackage{graphicx}
\usepackage{lscape}
\usepackage{subfigure}
\usepackage{natbib}
\usepackage{hyperref}
\usepackage[varg]{txfonts}
\usepackage{longtable}
\usepackage{booktabs}
\usepackage[normalem]{ulem} 
 \usepackage{float}
\usepackage{multirow}
\usepackage{soul}
\usepackage[table,x11names]{xcolor}
\usepackage{scalerel}
\usepackage{tabularx}
\usepackage{tikz}
\usepackage{ulem}
\definecolor{vpc}{rgb}{0.2, 0.8, 0.2}

\newenvironment{Acknowledgements}{\section*{Acknowledgements}}{}

\usetikzlibrary{svg.path}

\definecolor{orcidlogocol}{HTML}{A6CE39}
\tikzset{
  orcidlogo/.pic={
    \fill[orcidlogocol] svg{M256,128c0,70.7-57.3,128-128,128C57.3,256,0,198.7,0,128C0,57.3,57.3,0,128,0C198.7,0,256,57.3,256,128z};
    \fill[white] svg{M86.3,186.2H70.9V79.1h15.4v48.4V186.2z}
                 svg{M108.9,79.1h41.6c39.6,0,57,28.3,57,53.6c0,27.5-21.5,53.6-56.8,53.6h-41.8V79.1z M124.3,172.4h24.5c34.9,0,42.9-26.5,42.9-39.7c0-21.5-13.7-39.7-43.7-39.7h-23.7V172.4z}
                 svg{M88.7,56.8c0,5.5-4.5,10.1-10.1,10.1c-5.6,0-10.1-4.6-10.1-10.1c0-5.6,4.5-10.1,10.1-10.1C84.2,46.7,88.7,51.3,88.7,56.8z};
  }
}

\newcommand\orcidicon[1]{\href{https://orcid.org/#1}{\mbox{\scalerel*{
\begin{tikzpicture}[yscale=-1,transform shape]
\pic{orcidlogo};
\end{tikzpicture}
}{|}}}}
\bibpunct{(}{)}{;}{a}{}{,} 

\usepackage{amstext}
\usepackage{amsmath}
\newcommand{\FeatureImportancePlot}[1]{%
    \begin{tikzpicture}
        \def\plotwidth{4.85}
        
        \pgfmathsetmacro{\maxvalue}{0}
        \foreach \feature/\importance/\error in {#1} {
            \pgfmathsetmacro{\temp}{\importance + \error}
            \ifdim \temp pt > \maxvalue pt \global\let\maxvalue\temp \fi
        }
        
        \pgfmathsetmacro{\scaleFactor}{\plotwidth / \maxvalue}
        
        \def\barheight{0.3}
        \def\verticalstep{0.4}
        
        \pgfmathsetmacro{\featurecount}{0}
        \foreach \feature/\importance/\error in {#1} {
            \pgfmathsetmacro{\featurecount}{\featurecount + 1}
        }
        
        \foreach \feature/\importance/\error [count=\i] in {#1} {
            \pgfmathsetmacro{\ypos}{-(\i-1)*\verticalstep}
            \pgfmathsetmacro{\scaledImportance}{\importance * \scaleFactor}
            \pgfmathsetmacro{\scaledError}{\error * \scaleFactor}
            
            \fill[blue!50] (0,\ypos) rectangle (\scaledImportance, \ypos+\barheight);
            
            \draw[red, very thick] 
                (\scaledImportance-\scaledError, \ypos+\barheight/2) -- 
                (\scaledImportance+\scaledError, \ypos+\barheight/2);
            \draw[red, very thick] 
                (\scaledImportance-\scaledError, \ypos+\barheight/4) -- 
                (\scaledImportance-\scaledError, \ypos+3*\barheight/4);
            \draw[red, very thick] 
                (\scaledImportance+\scaledError, \ypos+\barheight/4) -- 
                (\scaledImportance+\scaledError, \ypos+3*\barheight/4);
            
            \node[anchor=east, font=\small] at (-0.2,\ypos+\barheight/2) {\feature};
        }
        
        \pgfmathsetmacro{\xaxispos}{0.7}
        \draw[->] (0,\xaxispos) -- (\plotwidth,\xaxispos);

         \foreach \x in {0,0.2,0.4,0.6,0.8,1} {
            \pgfmathsetmacro{\labelx}{\x * \plotwidth}
            \pgfmathsetmacro{\truevalue}{\x * \maxvalue}
            \draw[thin, gray!50] (\labelx,\xaxispos-0.1) -- (\labelx,\xaxispos+0.1);
            \node[below, font=\footnotesize] at (\labelx,\xaxispos+0.5) {\pgfmathprintnumber[fixed, precision=2]{\truevalue}};
        }
        
        \node[font=\small] at ({\plotwidth/2},\xaxispos+0.7) {\textbf{Feature Importance}};
    \end{tikzpicture}
}

\hypersetup{
    colorlinks=true,
    citecolor=blue,
    linkcolor=blue,
    urlcolor=blue,
    }

\makeatletter
\renewcommand*\aa@pageof{, page \thepage{} of \pageref*{LastPage}}
\makeatother
\definecolor{imputedR}{rgb}{0.99608, 0.87843, 0.56471}
\definecolor{imputedG}{rgb}{1, 1, 0.74902}
\definecolor{both}{rgb}{0.56863, 0.74902, 0.85882}
\definecolor{cornflowerblue}{rgb}{0.39, 0.58, 0.93}
\definecolor{ste}{rgb}{0.99, 0.00, 0.5}

\begin{document}

   \title{Selection of optically variable active galactic nuclei via a random forest algorithm\thanks{Observations were provided by the ESO programs 088.D-4013, 092.D-0370, and 094.D-0417 (PI G. Pignata).}}

   \author{D. De Cicco\inst{1,2,3,\orcidicon{0000-0001-7208-5101}}, G. Zazzaro\inst{4,\orcidicon{0000-0001-6042-6650}}, S. Cavuoti\inst{3,5,\orcidicon{0000-0002-3787-4196}}, M. Paolillo\inst{1,3,5,\orcidicon{0000-0003-4210-7693}}, G. Longo\inst{1,\orcidicon{0000-0002-9182-8414}}, 
   V. Petrecca\inst{1,3\orcidicon{0000-0002-3078-856X}}, I. Saccheo\inst{6,7\orcidicon{0000-0003-1174-6978}}, P.~Sánchez-Sáez\inst{8,2,\orcidicon{0000-0003-0820-4692}}}
   \authorrunning{D. De Cicco et al.}

\institute{
Department of Physics, University of Napoli ``Federico II'', via Cinthia 9, 80126 Napoli, Italy
\\e-mail: demetra.decicco@unina.it
\and
Millennium Institute of Astrophysics (MAS), Nuncio Monse\~nor Sotero Sanz 100, Providencia, Santiago, Chile    
\and
INAF - Osservatorio Astronomico di Capodimonte, via Moiariello 16, 80131 Napoli, Italy 
\and
CIRA - Centro Italiano di Ricerche Aerospaziali, via Maiorise s.n.c., 81043 Capua, Italy 
\and
INFN - Sezione di Napoli, via Cinthia 9, 80126 Napoli, Italy 
\and
Dipartimento di Matematica e Fisica, Università Roma Tre, Via 
della Vasca Navale 84, 00146 Roma, Italy
\and
INAF - Osservatorio astronomico di Roma, Via Frascati 33, I-00040
Monte Porzio Catone, Italy                                     
\and
European Southern Observatory, Karl-Schwarzschild-Strasse 2, 85748 Garching bei M\"{u}nchen, Germany 
  }

\date{}%Received September 1, 2023; accepted March 16, 1997
% \abstract{}{}{}{}{} 
% 5 {} token are mandatory
  \abstract
   {
   A defining characteristic of active galactic nuclei (AGN) that distinguishes them from other astronomical sources is their stochastic variability, which is observable across the entire electromagnetic spectrum. Upcoming optical wide-field surveys, such as the Vera C. Rubin Observatory's Legacy Survey of Space and Time, are set to transform astronomy by delivering unprecedented volumes of data for time domain studies. This data influx will require the development of the expertise and methodologies necessary to manage and analyze it effectively.
   }
   {
   This project focuses on optimizing AGN selection through optical variability in wide-field surveys and aims to reduce the bias against obscured AGN. We tested a random forest (RF) algorithm trained on various feature sets to select AGN. The initial dataset consisted of 54 observations in the $r$-band and 25 in the $g$-band of the COSMOS field, captured with the VLT Survey Telescope over a 3.3-year baseline.
   }
   {
   Our analysis relies on feature sets derived separately from either band plus a set of features combining data from both bands, mostly characterizing AGN on the basis of their variability properties and obtained from their light curves. We trained multiple RF classifiers using different subsets of selected features and assessed their performance via targeted metrics.
    }
   {   
   Our tests provide valuable insights into the use of multiband and multivisit data for AGN identification. We compared our findings with previous studies and dedicated part of the analysis to potential enhancements in selecting obscured AGN. The expertise gained and the methodologies developed here are readily applicable to datasets from other ground- and space-based missions.
   }
   {}


   \maketitle

\section{Introduction}
\label{section:introduction}
Active galactic nuclei (AGN) are undoubtedly among the most fascinating sources in the Universe. They are powered by central supermassive black holes (e.g., \citealt{BBS,Salpeter64,Rees84,KR95}, and references therein) and display diverse characteristics across different wavebands due to both observational perspectives and intrinsic properties. A straightforward consequence of such differences is that no identification technique is capable of returning a complete sample of AGN, as almost a century of investigation has made clear. Nevertheless, AGN do share a key property, this being variability in both their continuum and line emission (though the amplitude and timescales of this variability differ depending, once again, on the waveband). In virtue of this, AGN selection through variability in optical and infrared (IR) bands offers a unique advantage, as it can leverage the multi-visit surveys from ground-based observatories, which have accumulated over the past few decades \citep[e.g.,][]{Ulrich93,Trevese89,Trevese94,Trevese,Klesman&Sarajedini,sarajedini11,decicco15,decicco19,sanchezsaez,PSS23}. New-generation telescopes -- such as the Vera C. Rubin Observatory \citep{ivezich19}, expected to begin operations in mid-2025 -- promise to revolutionize time-domain astronomy by providing the astronomical community with a groundbreaking data volume that will reshape our understanding of the Universe.

This work is part of a series dedicated to the search for AGN in the COSMOS field based on their optical variability and centered on the analysis of time series from the VLT Survey Telescope (VST). The VST is a 2.6-meter optical telescope located at the Paranal Observatory in Chile and designed for wide-field imaging surveys of the southern sky, with a field of view of 1 sq. deg. and a pixel scale of 0.214\arcsec \citep{VST}. Specifically, here we test the use of a random forest (RF; \citealt{Breiman2001}) algorithm to select AGN on the basis of different sets of features quantifying their variability. Over the past decade, the VST time series have been extensively utilized to explore a range of topics, such as variable stars, transient events, and cosmology~\citep[e.g.,][]{Cappellaro15,Falocco15,decicco15,decicco19,decicco21,Botticella17,Fu18,Liu18,Liu20,Poulain20}, as well as for more technical purposes, including the development of an outlier detection pipeline \citep{cavuoti24}. 
In particular, in the context of AGN selection, \citet{decicco21} is a precursor work in our series dedicated to the VST-COSMOS field, where we initially tested the performance of a model based on an RF algorithm for AGN identification through optical variability, examining how different labeled sets (LSs) and sets of features affect the selection. Most features were selected because they characterize variability and were derived from the source light curves, but we also included six color features and a morphology indicator. That work confirmed the well-known reliability of optical variability as a tool for identifying unobscured AGN, yielding slightly better results than those of \citet{decicco19}, where a traditional approach had been adopted for the selection of AGN. Respectively, we obtained 91\% versus 86\% precision\footnote{Precision is defined as the ratio of the sources correctly classified as AGN (true positives) to all the sources classified as AGN regardless of whether the classification is correct or not (true positives+false positives). Hence, it gives the fraction of AGN correctly classified.}; 69\% versus 59\% recall\footnote{Recall is defined as the ratio of the sources correctly classified as AGN (true positives) to all the known AGN in the LS regardless of whether they have been correctly classified or not (true positives + false negatives). Hence, it reveals how often known AGN are correctly classified.} for the identification of spectroscopically confirmed AGN; 94\% versus 82\% recall for the identification of spectroscopically confirmed Type I AGN; 21\% versus 18\% recall for the identification of spectroscopically confirmed Type II AGN (see Table 5 in \citealt{decicco21} for additional information). 
This last result might appear discouraging, yet it should not be surprising at all considering that the optical emission from Type II AGN is dominated by the host galaxy contribution. Indeed, shorter baselines had yielded even poorer results: when the baseline was limited to five months, \citet{decicco15} found a 6\% recall for Type II AGN, in agreement with what is usually found in the literature.

Identifying Type II AGN through optical variability has notoriously posed challenges for decades, classically attributed to their different orientation (e.g., \citealt{Antonucci,Urry&Padovani}; but see also \citealt{lamassa15,macleod16,Green22}, and references therein for a more detailed insight of this class of sources). Indeed, as they are observed roughly edge-on, the dust structure surrounding the accretion disk often prevents one from observing the disk emission, which primarily peaks in the UV/optical waveband.

In essence, this study is meant to expand on the AGN selection approach via RF presented in \citet{decicco21}, taking advantage of the quality and cadence of the VST-COSMOS survey but making use of a significantly larger feature set. Indeed, in our previous work, the variability features were derived solely from the $r$-band, which has the highest observing cadence among the VST-COSMOS bands. Here we start with those same $r$-band features but extend the set by adding the corresponding features computed from $g$-band data and by testing the inclusion of bivariate features, that is to say, features that combine data from both the $r$ and $g$ bands. Such features are expected to improve the selection, as many multiwavelength monitoring campaigns of AGN show correlations between variability properties in adjacent regions of the electromagnetic spectrum \citep[e.g.,][]{Edelson,VandenBerk}.
We test several RF classifiers that always operate on the same LS but use different feature sets for each test. Because our $r-$ and $g-$band data were not always obtained simultaneously, we resort to data imputation in order to obtain, for each source, light curves covering the same baseline and with an equivalent number of simultaneous observations in the two bands used.

This work serves as a forecasting study in view of the highly anticipated Legacy Survey of Space and Time \citep[e.g.,][]{lsst} from the already mentioned Rubin Observatory, and our series of tests are aimed at evaluating whether the use of multiband data enhances AGN selection and at examining the impact of the observing cadence as well as the inclusion of synthetic visits on selection accuracy. We also dedicate part of our analysis to the optimization of the selection of obscured AGN. 

The structure of this paper is as follows: Section \ref{section:dataset} introduces the dataset and the class imbalance problem affecting our LS. Section \ref{section:data_preparation} describes the data imputation process and the set of features used and illustrates how we train our classifiers. Section \ref{section:rf_tests} presents the various classifiers tested, with a focus on the selection of obscured AGN. Section \ref{section:conclusions} summarizes our main findings.

We note that throughout this work we tend to prefer the definitions ``unobscured'' and ``obscured'' -- to stress the relative dominance of the nucleus on the host galaxy -- rather than the corresponding terms ``Type I'' and ``Type II''. Indeed, while there is a correspondence between these two pairs of labels for AGN, we consider the distinction in types too strict and static, especially in light of various studies conducted in recent years, which have shown how AGN can ``change'' their type and exist as something between Type I and Type II (see references above). Nevertheless, in this work, we use catalogs of AGN selected by \citet{marchesi} as Type I and Type II based on their spectral properties.

\section{Data understanding} 
\label{section:DU}
The data understanding phase involved a detailed exploration of the dataset. It emphasized the extraction of domain-relevant insights and the identification of patterns, anomalies, or limitations that may influence downstream analyses or model development.

\subsection{The VST-COSMOS dataset}
\label{section:dataset}
This work is based on a series of 54 $r$-band and 25 $g$-band visits of the COSMOS field captured by the VST, obtained during three observing seasons and spanning the same baseline of 3.3~yr. Each visit originally covered a $\approx1$ sq. deg. area, but we masked $\approx17\%$ of it (mostly edges, defected areas, and saturated stars). The $r$-band dataset has been widely used in previous studies dedicated to AGN optical variability \citep{decicco15,decicco19,decicco21,decicco22}, while the $g$-band dataset is used for the first time for this kind of study. During the first observing seasons, corresponding to the first five months of observations, the planned observing cadence for the $r$ and $g$ band was of $\approx$ 3 and $\approx$~10 days, respectively (with several observing constraints that affected both), which explains the differences in the sampling for the two bands. This led us to resort to data imputation, essentially consisting of replacing missing or incomplete data within a dataset to minimize the biases and errors arising from the lack of original data, as detailed in Sect. \ref{section:data_imputation}. This allowed us to use bivariate features at the expense of some arbitrariness in the imputation method.

The baseline of this dataset is currently being extended to more than 11 yr with two additional observing seasons, and it will be even longer, as more observations are ongoing. The visit depth is $r \lesssim 24.6$ and $g \lesssim 24.2$ mag for point sources (${\sim}5\sigma$ confidence level). This dataset can therefore be considered as a scaled-down version of what will be obtained from the LSST main survey, which is expected to cover a 10 yr baseline, with single-visit depths of 24.7 and 25.0 mag in the $r$ and $g$ bands, respectively. This is one of the reasons why in this series of works we have been extensively exploiting our dataset for LSST performance forecasting studies. 

We report essential information about our dataset in Table~\ref{tab:dataset}, while we refer the reader to the above-mentioned papers for further details about the $r$-band dataset, and in particular we refer to \citet{decicco15} for an overview of the reduction process. Throughout this work, magnitudes are in the AB system and time differences are expressed in the observer reference frame. 

\begin{table*}[htb]
\caption{\footnotesize{VST-COSMOS dataset for the $r$ and $g$ bands.}} 
\label{tab:dataset}      
\begin{minipage}{0.495\textwidth}
\begin{tabular}{c c c c c}
\toprule
ID & obs. date & time (days) & $r$-band & $g$-band\\
\midrule
1 & 2011-Dec-18 & 0 & y\phantom{*} & n\phantom{*}\\
2 & 2011-Dec-22 & 4 & y\phantom{*} & n\phantom{*}\\
\rowcolor{both}3 & 2011-Dec-27 & 9 & y\phantom{*} & y\phantom{*}\\
4 & 2011-Dec-31 & 13 & y\phantom{*} & n\phantom{*}\\
5 & 2012-Jan-02 & 15 & y\phantom{*} & n\phantom{*}\\
6 & 2012-Jan-06 & 19 & y\phantom{*} & n\phantom{*}\\ 
7 & 2012-Jan-18 & 31 & y\phantom{*} & n\phantom{*}\\
8 & 2012-Jan-20 & 33 & y\phantom{*} & n\phantom{*}\\	
\rowcolor{both}9 & 2012-Jan-22 & 35 & y\phantom{*} & y\phantom{*}\\	
\rowcolor{imputedG}10 & 2012-Jan-24 & 37 & y\phantom{*} & n*\\
\rowcolor{imputedG}11 & 2012-Jan-27 & 40 & y\phantom{*} & n*\\
\rowcolor{imputedG}12 & 2012-Jan-29 & 42 & y\phantom{*} & n*\\
\rowcolor{both}13 & 2012-Feb-02 & 46 & y\phantom{*} & y\phantom{*}\\
\rowcolor{both}14 & 2012-Feb-16 & 60 & y\phantom{*} & y\phantom{*}\\
\rowcolor{imputedG}15 & 2012-Feb-19 & 63 & y\phantom{*} & n*\\
\rowcolor{imputedG}16 & 2012-Feb-21 & 65 & y\phantom{*} & n*\\
\rowcolor{imputedG}17 & 2012-Feb-23 & 67 & y\phantom{*} & n*\\
\rowcolor{both}18 & 2012-Feb-26 & 70 & y\phantom{*} & y\phantom{*}\\
19 & 2012-Feb-29 & 73 & y\phantom{*} & n\phantom{*}\\
20 & 2012-Mar-03 & 76 & y\phantom{*} & n\phantom{*}\\
21 & 2012-Mar-13 & 86 & y\phantom{*} & n\phantom{*}\\
22 & 2012-Mar-15 & 88 & y\phantom{*} & n\phantom{*}\\
\rowcolor{both}23 & 2012-Mar-17 & 90 & y\phantom{*} & y\phantom{*}\\
24 & 2012-May-08 & 142 & y\phantom{*} & n\phantom{*}\\	
\rowcolor{imputedR}25 & 2012-May-09 & 143 & n* & y\phantom{*}\\
26 & 2012-May-11 & 145 & y\phantom{*} & n\phantom{*}\\	
27 & 2012-May-17 & 151 & y\phantom{*} & n\phantom{*}\\	
28 & 2013-Dec-15 & 728 & n\phantom{*} & y\phantom{*}\\
29 & 2013-Dec-27 & 740 & y\phantom{*} & n\phantom{*}\\	
30 & 2013-Dec-30 & 743 & y\phantom{*} & n\phantom{*}\\	
\rowcolor{both}31 & 2014-Jan-03 & 747 & y\phantom{*} & y\phantom{*}\\	
32 & 2014-Jan-05 & 749 & y\phantom{*} & n\phantom{*}\\	
33 & 2014-Jan-12 & 756 & y\phantom{*} & n\phantom{*}\\	
\bottomrule
\end{tabular}
\end{minipage} \hfill
\begin{minipage}{0.495\textwidth}
\begin{tabular}{c c c c c}
\toprule
ID & obs. date & time (days) & $r$-band & $g$-band\\
\midrule
\rowcolor{both}34 & 2014-Jan-21 & 765 & y\phantom{*} & y\phantom{*}\\
35 & 2014-Jan-24 & 768 & y\phantom{*} & n\phantom{*}\\	
36 & 2014-Feb-09 & 784 & y\phantom{*} & n\phantom{*}\\
\rowcolor{both}37 & 2014-Feb-19 & 794 & y\phantom{*} & y\phantom{*}\\	
\rowcolor{imputedG}38 & 2014-Feb-21 & 796 & y\phantom{*} & n*\\	
\rowcolor{imputedG}39 & 2014-Feb-23 & 798 & y\phantom{*} & n*\\	
\rowcolor{both}40 & 2014-Feb-26 & 801 & y\phantom{*} & y\phantom{*}\\	
\rowcolor{imputedG}41 & 2014-Feb-28 & 803 & y\phantom{*} & n*\\	
\rowcolor{imputedR}42 & 2014-Mar-04 & 807 & n* & y\phantom{*}\\
43 & 2014-Mar-08 & 811 & y\phantom{*} & n\phantom{*}\\	
\rowcolor{both}44 & 2014-Mar-21 & 824 & y\phantom{*} & y\phantom{*}\\	
\rowcolor{imputedG}45 & 2014-Mar-23 & 826 & y\phantom{*} & n*\\	
\rowcolor{imputedG}46 & 2014-Mar-25 & 828 & y\phantom{*} & n*\\
\rowcolor{both}47 & 2014-Mar-29 & 832 & y\phantom{*} & y\phantom{*}\\	
\rowcolor{both}48 & 2014-Apr-04 & 838 & y\phantom{*} & y\phantom{*}\\	
49 & 2014-Apr-07 & 841 & y\phantom{*} & n\phantom{*}\\	
50 & 2014-Dec-03 & 1081 & y\phantom{*} & n\phantom{*}\\	
51 & 2014-Dec-16 & 1094 & n\phantom{*} & y\phantom{*}\\
52 & 2014-Dec-25 & 1103 & n\phantom{*} & y\phantom{*}\\
53 & 2015-Jan-03 & 1112 & n\phantom{*} & y\phantom{*}\\
\rowcolor{imputedG}54 & 2015-Jan-10 & 1119 & y\phantom{*} & n*\\	
55 & 2015-Jan-15 & 1124 & n\phantom{*} & y\phantom{*}\\
56 & 2015-Jan-23 & 1132 & n\phantom{*} & y\phantom{*}\\
\rowcolor{imputedG}57 & 2015-Jan-28 & 1137 & y\phantom{*} & n*\\	
\rowcolor{imputedR}58 & 2015-Jan-30 & 1139 & n* & y\phantom{*}\\
\rowcolor{imputedG}59 & 2015-Jan-31 & 1140 & y\phantom{*} & n*\\
60 & 2-15-Feb-14 & 1154 & n\phantom{*} & y\phantom{*}\\
61 & 2015-Feb-15 & 1155 & y\phantom{*} & n\phantom{*}\\	
62 & 2015-Mar-10 & 1178 & y\phantom{*} & n\phantom{*}\\	
\rowcolor{imputedR}63 & 2015-Mar-13 & 1181 & n* & y\phantom{*}\\
\rowcolor{imputedG}64 & 2015-Mar-14 & 1182 & y\phantom{*} & n*\\	
\rowcolor{imputedG}65 & 2015-Mar-19 & 1187 & y\phantom{*} & n*\\	
66 & 2015-Mar-22 & 1190 & n\phantom{*} & y\phantom{*}\\
\bottomrule
\end{tabular}
\end{minipage} \hfill
\vspace{1mm}
\footnotesize{\textbf{Notes.} The table reports the visit ID, date, time in days from the first observation, presence or absence (y/n) of a visit in the $r$ and $g$ bands. Different colors are associated to different y/n combinations for the two bands. The asterisk identifies the visits that were originally missing in either band, and that we included in the dataset via data imputation, as detailed in Sect. \ref{section:data_imputation}.}
\end{table*}

\subsection{The labeled set and the class imbalance problem}
\label{section:icp}
One of the issues one commonly has to face when handling real data is the imbalance in the number of objects of different types. Specifically, for this study we had to deal with an unbalanced LS, consisting of two main classes: 380 AGN and 2163 non-AGN, of which 1,168 are stars and 995 are ``inactive'' galaxies, meaning galaxies where no nuclear activity is detected. The sources that make up our LS were selected from different catalogs from the literature, essentially adopting the same criteria described in \citet{decicco21}. In short, stars come from the COSMOS ACS catalog \citep{Koekemoer,scoville07b} and the selection was refined by checking that these sources lie on the stellar locus on a $r-z$ versus $z-K$ diagram\footnote{As mentioned above, Sect. 2.5 of \citet{decicco21} explains in detail how we selected a sample of 1,000 stars. Here we simply did not set the size of the star LS to 1,000 a priori; furthermore, contrary to what was done in \citet{decicco21}, here we included in the star LS six sources with $r-z > 1.5$ mag, as their classification as stars seems to be reliable.} \citep{nakos}. Inactive galaxies were selected from the COSMOS2015 catalog \citep{laigle}, which provides a classification based on the best-fit templates from \citet{bc03}. We cross-matched our samples of stars and inactive galaxies with COSMOS catalogs available from the literature and excluded sources with conflicting classifications. Additional details can be found in Sect. 2.5 of \citet{decicco21}. For what concerns AGN, the LS here used is smaller than the one used in \citet{decicco21}, which was based on $r$-band data only, as here we had to take into account the $g$ band as well; as a consequence, here we excluded all the sources that were not detected in at least two visits in the $g$ band, that is, the minimum required to compute the various variability features used. This requirement generally affects the non-AGN LS as well, but in the present work we compensated for the exclusion of some stars by including some others, while the size of the inactive galaxy LS is not significantly reduced by the above-mentioned requirement (only five sources are excluded from the present LS). 

As in \citet{decicco21}, we identify AGN on the basis of different diagnostics, and classify them as spectroscopic Type I (i.e., unobscured, 217 sources) and Type II (i.e., obscured, 104), and mid-infrared-selected (MIR, 211, 59 of which lack spectroscopic classification). Specifically, the spectroscopic classification for Type I and Type II AGN comes from the \emph{Chandra}-COSMOS Legacy Catalog \citep{marchesi}, which means these are X-ray emitting sources for which an optical counterpart is available and that were classified as AGN via optical spectroscopy, based on the traditional criterion relying on the presence of broad ($\geq 2,000$~km s$^{-1}$) emission lines in a spectrum; the MIR classification is based on the criterion by \citet{donley} which, in a diagram comparing the two MIR colors $\log(F[8.0]\mu\mbox{m}/F[4.5]\mu\mbox{m})$ versus $\log(F[5.8]\mu\mbox{m}/F[3.6]\mu\mbox{m})$, identifies a region where AGN typically place themselves, the MIR information coming from the already mentioned COSMOS2015 catalog; see also Sect. 4 of \citealt{decicco19} for further details. We note that a source can be classified as AGN by more than one criterion. In addition, we note that we also have information about the nature of the AGN in our LS, based on their X-ray emission \citep{marchesi,brusa} and on their optical variability; nevertheless, we decided not to use such information in this work, as we are interested in a proper comparison with what we did in \citet{decicco21}, where we only consider spectroscopic- and MIR-selected AGN for our LS. 
We also note that while we always know which source in the non-AGN LS is a star and which one is a galaxy, for our purpose they all are simply considered as non-AGN. Hence, our classification will be binary, AGN being the minority (or positive) class, and non-AGN being the majority (or negative) class.

\begin{figure*}[ht!]
 \centering
\subfigure
            {\includegraphics[width=\columnwidth]{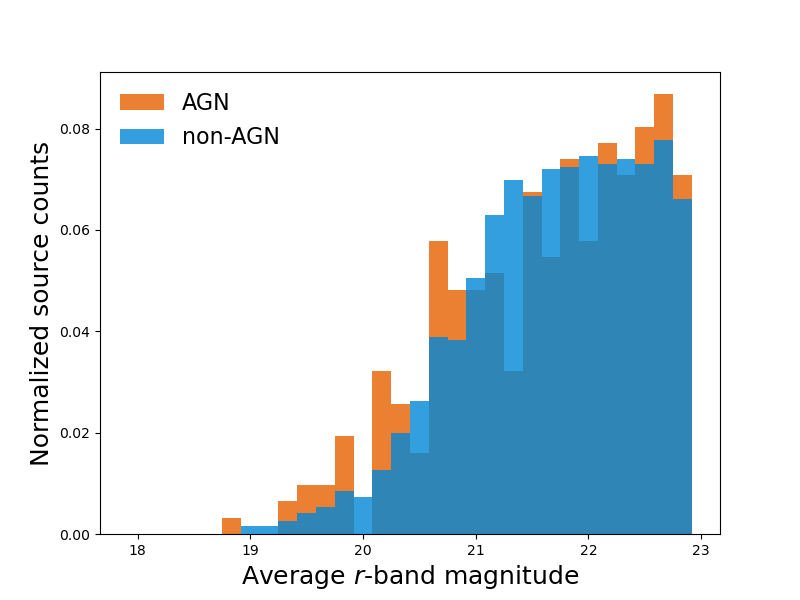}}
\subfigure
            {\includegraphics[width=\columnwidth]{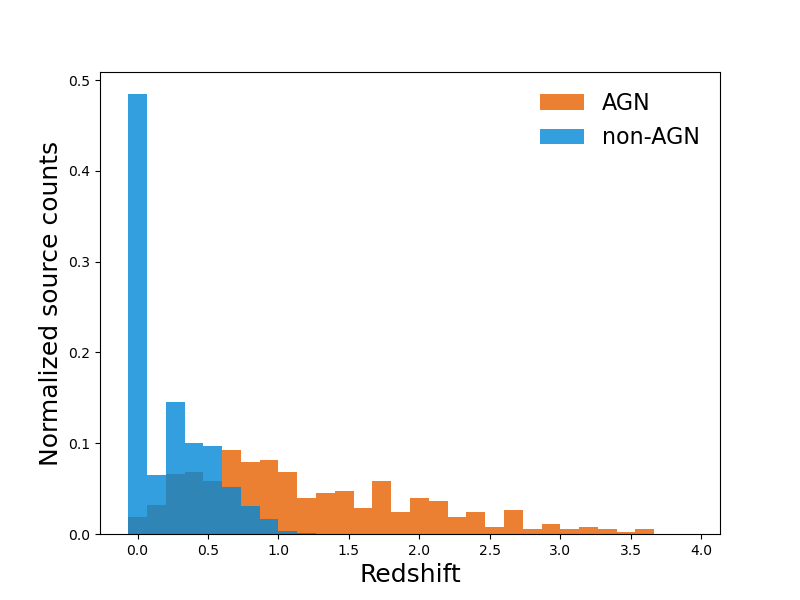}}
\subfigure
            {\includegraphics[width=\columnwidth]{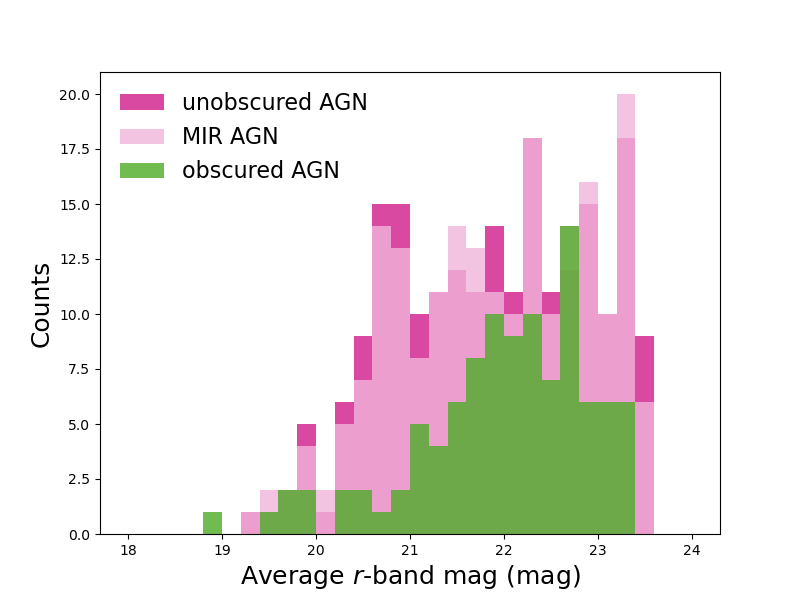}}
\subfigure
            {\includegraphics[width=\columnwidth]{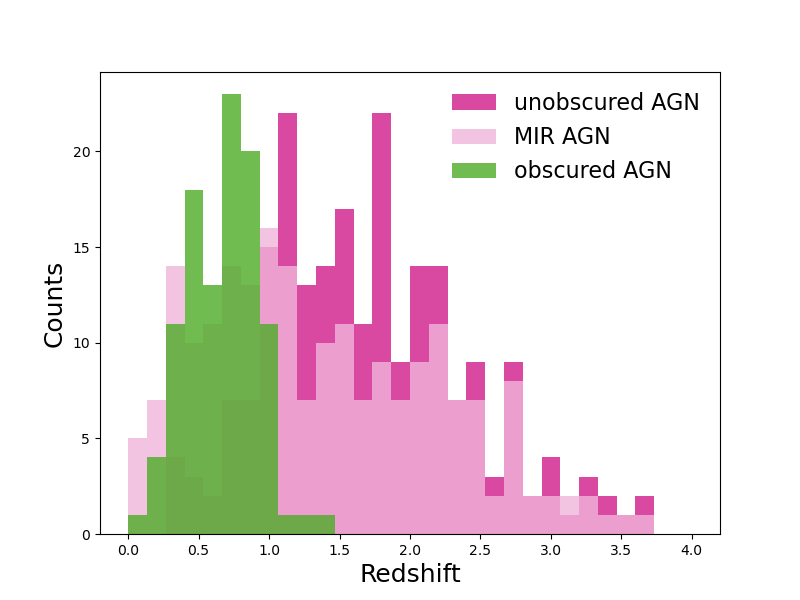}}
  \caption{\footnotesize{Average $r$-band magnitude (\textit{upper-left panel}) and redshift (\textit{upper-right panel}) for the two classes (AGN and non-AGN) in our LS. Each histogram has been normalized to the total number of sources in the corresponding class. Average $r$-band magnitude (\textit{lower-left panel}) and redshift (\textit{lower-right panel}) for the AGN LS and the three subsamples of unobscured, obscured, and MIR AGN.}}\label{fig:hists}
   \end{figure*}

In Fig. \ref{fig:hists} we show the magnitude and redshift distributions for the two classes of AGN and non-AGN in our LS, and also for the three subclasses that form our AGN LS. Consistent with previous works, we cut our LS to an average magnitude $r \leq 23.5$ mag, which is assumed as the completeness limit of our survey. For what concerns redshifts, the median value is 0.411 for the galaxies in the non-AGN class, being the redshift 0 for the stars, and 1.106 for AGN, which extend to higher redshifts, the highest being 3.715. If we focus on unobscured, obscured, and MIR-selected AGN, we can see that their magnitudes roughly span the same range, while obscured AGN have lower redshifts than the other two subclasses, the median values being 1.657, 0.678, and 1.214 for unobscured, obscured, and MIR AGN, respectively. We stress that, while the first two subclasses are disjointed, the MIR subclass partly overlaps the other two.

The classifier training phase is often deeply conditioned by the majority class: though the models can have high general accuracy, at a closer look they show a low predictive accuracy for the minority class. Models trained with most learning algorithms on an unbalanced dataset frequently predict most records as negative. This is often regarded as a problem in learning from highly imbalanced datasets, especially when the minority class is the one we are interested in.

Imbalance is quantified by the imbalance ratio \citep[e.g.,][]{amin}, defined as the ratio of minority class instances to majority ones (also referred to as skew; e.g., \citealt{Jeni}). Based on the above-listed numbers, the imbalance ratio of our LS, corresponding to the ratio between AGN and non-AGN, is 380/2163 = 0.1757. Intuitively, the greater the imbalance in a dataset, the more complex the learning process. Hence the training of a classifier with satisfactory performance is increasingly challenging.

\section{Data preparation}
\label{section:data_preparation}
In the present section we illustrate how we dealt with the different sampling cadences of the two bands we used in this work. We also introduce the features here used, focusing on the ones that are new with respect to those used in \citet{decicco21}.

\subsection{Missing data imputation}
\label{section:data_imputation}
Ideally, in order to compute a feature based on data from two bands, we should have ``simultaneous'' -- meaning as close as possible in time -- visits in these two bands. Hence, we considered as good candidates for the computation of bivariate features only those pairs of visits that were obtained during the same night. This condition is fulfilled by 13 out of the 66 observing dates in our list; it is also worth noting that in four instances there is a one-night lag between observations in the $r$ and $g$ bands. Although ideally we would like to replicate the LSST observing cadence, where observations in different bands will be obtained in the same night, this shift of one night should not be a major issue for all the non-blazar objects. Based on this, with our dataset we should be able to build bivariate features using only 13 points per light curve in the best case scenario, that is, when a source is detected in each of the 13 corresponding visits. This would mean that most of our dataset would be wasted. Also, since the differences in the sampling of the two bands lead to a different number of visits for each of them, the various features that we would compute for each source would be obtained from a different number of visits. Hence, the weight of the various features in any rankings would be different as it would depend on the number of visits involved in their computation. We therefore resorted to data imputation in order to fill the gaps in the light curves when possible, as detailed in the following. We compared the data in the $g$ and $r$ bands, and considered all the cases where, on a given date, there is a visit in either band, but not in both. Hence we considered the closest visits before and after the given date in the band with the gap, and computed the time difference between the two, in days: if this is $\leq 15$ days, we considered this as a good case for data imputation, that is, we assume that, knowing the properties of AGN variability, variations in such a time interval will not be too distant from a linear behavior. We can therefore fill the gap in the dataset via linear interpolation. We did this for both bands, filling four gaps in the $r$ band, that is, the one with the denser sampling, and 16 gaps in the $g$ band. We did not perform imputation when the time difference is $> 15$ days. After this procedure we ended up with 33 visits per band. The error bars we associated to both real and synthetic magnitude values were computed from the whole sample of VST-COSMOS sources: we considered the average magnitude value for each source, and defined the error bar as the 95\% uncertainty on that magnitude value.

\subsection{Features used for AGN selection}
\label{section:features}
The identification of AGN in this work is based on the use of a number of features. Most of them have been frequently used in the literature for variability studies, as they are suitable for variability analysis and can be computed from the source light curves; these features, which hereafter we refer to as ``univariate'', were computed independently for the $r$ and the $g$ band from the corresponding light curves. Specifically, we used the same univariate features used in \citet{decicco21}. Following that work, we also used the same set of color features as well as the only morphology feature there used; of course these were computed just once, being independent on our $g-$ and $r-$band light curves. Details about these features can be found in Sects. 2.2, 2.3, and 2.4 of \citet{decicco21}, but we report here Table 1 from that work -- Table \ref{tab:features} in this work -- for the sake of convenience. In addition to the above-mentioned features, in this work we included a set of features that were computed combining the $g$ and $r$ bands together and that we therefore labeled ``bivariate''. 
For what concerns the bivariate features we basically resorted to a series of similarity measures, quantifying the distance between the two light curves in each pair. Some of these measures, such as the $L_1$-norm or $L_2$-norm, are very well-known and of immediate interpretation, while some others are more complex. In general, these features are grouped under different families \citep[e.g.,][]{dist_meas}, namely:
\begin{itemize}
\item The $L_p$ Minkowski family, containing a series of measures corresponding to the generalized formula $\sqrt[p]{\sum_{i}|X_i-Y_i|^p}$ as the index $p$ changes; 
\item  The $L_1$ family, containing  measures related to the absolute difference $\sum_{i}|X_i-Y_i|$ introduced in the $L_1$ distance;
\item The intersection family, which  contains measures related to the intersection of the $g$ and $r$ sets of points;
\item The inner product family, where measures are defined on the basis of the inner product between the two sets of points in the two bands;
\item The fidelity family, where measures are defined starting from the so-called Fidelity similarity, that is, the sum of the square root of the inner product;
\item The $\chi^2$ family, whose member features originate from the square of the Euclidean norm $L_2$;
\item The Shannon entropy family, named after the Shannon entropy, which features in this group are based on;
\item The combination family, where characteristics from different families are combined;
\item The vicissitude family, which contains a number of features defined in \citet{Cha}.
\end{itemize}
The complete list of the bivariate features used in this work is reported in Table \ref{tab:bivariate}. 

In total, we used a set of: 2 $\times$ 29 univariate features + one morphological feature + 6 color features + 39 bivariate features + 29 features defined as the differences between homologous univariate $g-$ and $r-$band features + 29 features obtained from the $g-r$ light curves, for a total of 162 features. 

\begin{table*}[tb]
\caption{List of univariate variability features, morphology feature, and color features used in this work.}
\label{tab:features}      
 \renewcommand\arraystretch{1.2}
 \footnotesize
 \resizebox{\textwidth}{!}{
 \begin{tabular}{c l l l}
\toprule 
\ & Feature & Description & Reference\\
\hline
\multirow{9}{*}{{\rotatebox[origin=c]{90}{classic var. features}}} & \texttt{A$_{SF}$} & rms magnitude difference of the SF, computed over a 1 yr timescale & \citet{Schmidt}\\
& \ \texttt{$\gamma_{SF}$} & Logarithmic gradient of the mean change in magnitude & \citet{Schmidt}\\
& \ \texttt{GP\_DRW\_$\tau$} & Relaxation time $\tau$ (i.e., time necessary for the time series to become uncorrelated), & \citet{graham17}\\
& \  & from a DRW model for the light curve & \\
& \ \texttt{GP\_DRW\_$\sigma$} & Variability of the time series at short timescales ($t << \tau$), & \citet{graham17}\\
& \ & from a DRW model for the light curve & \\
& \ \texttt{ExcessVar} & Measure of the intrinsic variability amplitude & \citet{allevato}\\
& \ \texttt{P$_{var}$} & Probability that the source is intrinsically variable & \citet{mclaughlin}\\
& \ \texttt{IAR$_\phi$} & Level of autocorrelation using a  discrete-time representation of a DRW model & \citet{eyheramendy18}\\
\hline
\multirow{27}{*}{{\rotatebox[origin=c]{90}{FATS features}}} &\ \texttt{Amplitude} & Half of the difference between the median of the maximum 5\% and of the minimum & \citet{richards11}\\
 &\ & 5\% magnitudes & \\
 & \ \texttt{AndersonDarling} & Test of whether a sample of data comes from a population with a specific distribution & \citet{nun}\\
 & \ \texttt{Autocor\_length} & Lag value where the auto-correlation function becomes smaller than $\eta^e$ & \citet{kim11}\\
 & \ \texttt{Beyond1Std} & Percentage of points with photometric mag that lie beyond 1$\sigma$ from the mean & \citet{richards11}\\
 & \ \texttt{$\eta^e$} & Ratio of the mean of the squares of successive mag differences to the variance & \citet{kim14}\\
 & \ & of the light curve & \\
 & \ \texttt{Gskew} & Median-based measure of the skew & -\\
 & \ \texttt{LinearTrend} & Slope of a linear fit to the light curve & \citet{richards11}\\
 & \ \texttt{MaxSlope} & Maximum absolute magnitude slope between two consecutive observations & \citet{richards11}\\
 & \ \texttt{Meanvariance} & Ratio of the standard deviation to the mean magnitude & \citet{nun}\\
 & \ \texttt{MedianAbsDev} & Median discrepancy of the data from the median data & \citet{richards11}\\
 & \ \texttt{MedianBRP} & Fraction of photometric points within amplitude/10 of the median mag & \citet{richards11}\\
 & \ \texttt{MHAOV\_Period} & Periodo obtained via the Multi-Harmonic Analysis of Variability periodogram & \citet{Huijse18}\\
 & \ \texttt{PairSlopeTrend} & Fraction of increasing first differences minus fraction of decreasing first differences & \citet{richards11}\\
 & \ & over the last 30 time-sorted mag measures & \\
 & \ \texttt{PercentAmplitude} & Largest percentage difference between either max or min mag and median mag & \citet{richards11}\\
 & \ \texttt{Q31} & Difference between the 3\textsuperscript{rd} and the 1\textsuperscript{st} quartile of the light curve & \citet{kim14}\\
 & \ \texttt{Period\_fit} & False-alarm probability of the largest periodogram value obtained with LS & \citet{kim11}\\
 & \ \texttt{$\Psi_{CS}$} & Range of a cumulative sum applied to the phase-folded light curve & \citet{kim11}\\
 & \ \texttt{$\Psi_\eta$} & $\eta^e$ index calculated from the folded light curve & \citet{kim14}\\
 & \ \texttt{R$_{cs}$} & Range of a cumulative sum & \citet{kim11}\\
 & \ \texttt{Skew} & Skewness measure & \citet{richards11}\\
 & \ \texttt{Std} & Standard deviation of the light curve & \citet{nun}\\
 & \ \texttt{StetsonK} & Robust kurtosis measure & \citet{kim11}\\

\hline
\multirow{4}{*}{{\rotatebox[origin=c]{90}{morph.}}}\  \\
& \texttt{class\_star} & \emph{HST} stellarity index & \citet{Koekemoer},\\
\ & &  & \citet{scoville}\\ \\
\hline
\multirow{6}{*}{{\rotatebox[origin=c]{90}{colors}}} & \ \texttt{u-B} & CFHT $u$ magnitude -- Subaru $B$ magnitude & \citet{laigle}\\
 & \ \texttt{B-r} & Subaru SuprimeCam $B$ mag -- Subaru SuprimeCam $r$+ mag & \citet{laigle}\\
 & \ \texttt{r-i} & Subaru SuprimeCam $r+$ mag -- Subaru SuprimeCam $i+$ mag & \citet{laigle}\\
 & \ \texttt{i-z} & Subaru SuprimeCam $i+$ mag -- Subaru SuprimeCam $z++$ mag & \citet{laigle}\\
 & \ \texttt{z-y} & Subaru SuprimeCam $z++$ mag -- Subaru Hyper-SuprimeCam $y$ mag & \citet{laigle}\\
\cline{2-4}
 & \ \texttt{ch21} & \emph{Spitzer} 4.5 $\mu$m (\emph{channel2}) mag -- 3.6 $\mu$m (\emph{channel1}) mag & \citet{laigle}\\
\bottomrule
\vspace{1mm}
\end{tabular}
}
\footnotesize{\textbf{Notes.} The first two sections of the table report variability features; \texttt{class\_star} is the only morphology feature used; the bottom section of the table lists the color features used, where \texttt{ch21} is the only MIR color used, while the others are optical/near-infrared (NIR) colors. This table corresponds to Table 1 in \citet{decicco21}.}
\end{table*}

\begin{table*}[htb]
\caption{\footnotesize{List of the bivariate features used in this work grouped under the families introduced in Sect. \ref{section:features} \citep{dist_meas}.  
}} 
\label{tab:bivariate}   
\renewcommand{\arraystretch}{1.5}
\footnotesize
\subtable{
\begin{tabular}{l l}
\toprule
\ \textbf{Minkowski family} & \\
\midrule
\ City\_Block, $L_1$-norm & $d_{\mbox{\tiny{City}}}=\sum|g_i - r_i|$\\
\ Euclidean, $L_2$-norm & $d_{\mbox{\tiny{Eucl}}}=\sqrt{\sum(g_i - r_i)^2}$ \\
\ Chebyshev, $L_\infty$-norm & $d_{\mbox{\tiny{Cv}}}=\mbox{max}_i|g_i - r_i|$\\
\midrule
\ \textbf{$L_1$ family} & \\
\midrule
\ Sørensen & $d_{\mbox{\tiny{Sø}}}=\frac{\sum|g_i - r_i|}{\sum(g_i + r_i)}$\\
\ Gower & $d_{\mbox{\tiny{Gw}}}=\sum|g_i - r_i|/b$\\
\ Kulczynski & $d_{\mbox{\tiny{Kul}}}=\frac{\sum|g_i - r_i|}{\sum\min(g_i,r_i)}$\\
\ Canberra & $d_{\mbox{\tiny{Canb}}}=\sum{\frac{|g_i - r_i|}{(g_i + r_i)}}$\\
\ Lorentzian & $d_{\mbox{\tiny{Lor}}}=\sum\ln(1+|g_i - r_i|)$\\
\midrule
\ \textbf{Intersection family} & \\
\midrule
\ Intersection & $d_{\mbox{\tiny{Is}}}=\sum|g_i - r_i|/2$\\
\ Wave\_Hedges & $d_{\mbox{\tiny{WH}}}=\sum{\frac{|g_i - r_i|}{\max(g_i,r_i)}}$\\
\ Motyka & $s_{\mbox{\tiny{Mo}}}=\frac{\sum\max(g_i,r_i)}{\sum(g_i + r_i)}$\\
\ Czekanowski & $d_{\mbox{\tiny{Cz}}}=\frac{\sum|g_i-r_i|}{\sum(g_i + r_i)}$\\
\ Ruzicka & $s_{\mbox{\tiny{Mo}}}=\frac{\sum\min(g_i,r_i)}{\sum\max(g_i,r_i)}$\\
\midrule
\ \textbf{Inner product family} & \\
\midrule
\ Inner\_Product & $s_{\mbox{\tiny{Ip}}}=\sum g_i r_i$\\
\ Harmonic\_Mean & $s_{\mbox{\tiny{Hm}}}=2\frac{\sum g_i r_i}{(g_i + r_i)}$\\
\ Cosine & $s_{\mbox{\tiny{Cos}}}=\frac{\sum g_i r_i}{\sqrt{\sum g_i^2\sum r_i^2}}$\\
\ Jaccard & $d_{\mbox{\tiny{Ja}}}=\frac{\sum(g_i - r_i)^2}{\sum(g_i^2 + r_i^2 - g_i r_i)}$\\
\ Dice & $d_{\mbox{\tiny{Di}}}=\frac{\sum(g_i - r_i)^2}{\sum(g_i^2 + r_i^2)}$\\
\midrule
\ \textbf{Fidelity family} & \\
\midrule
\ Fidelity & $s_{\mbox{\tiny{Fid}}}=\sum\sqrt{g_i r_i}$\\
\ Bhattacharyya & $d_{\mbox{\tiny{Ba}}}=-\ln\sum\sqrt{g_i r_i}$\\
\ Squared-chord & $d_{\mbox{\tiny{SC}}}=\sum(\sqrt{g_i}-\sqrt{r_i})^2$\\
\bottomrule
\end{tabular}
}
\hfill
\renewcommand{\arraystretch}{1.5}
\footnotesize
        \subtable{
\begin{tabular}{l l}
\toprule
\textbf{$\chi^2$ family} \\
\midrule
\ Squared\_Euclidean & $d_{\mbox{\tiny{SE}}}=\sum(g_i - r_i)^2$\\
\ Pearson $\chi^2$ & $d_{\mbox{\tiny{Pea}}}=\sum(g_i - r_i)^2/r_i$\\
\ Neyman  $\chi^2$ & $d_{\mbox{\tiny{Ney}}}=\sum(g_i - r_i)^2/g_i$\\
\ $\chi^2$ & $d_{\mbox{\tiny{Sq$\chi$}}}=\sum\frac{(g_i - r_i)^2}{g_i + r_i}$\\
\ Divergence & $d_{\mbox{\tiny{Div}}}=2\sum\frac{(g_i - r_i)^2}{(g_i + r_i)^2}$\\
\ Clark & $d_{\mbox{\tiny{Cl}}}= \sqrt{\sum\left[\frac{|g_i - r_i|}{(g_i + r_i)}\right]^2}$\\
\ Additive\_Symmetric $\chi^2$ & $d_{\mbox{\tiny{Ad$\chi$}}}= \sum\frac{(g_i - r_i)^2(g_i + r_i)}{g_i r_i}$\\
\midrule
\textbf{Shannon's entropy family} \\
\midrule
\ Kullback-Leibler & $d_{\mbox{\tiny{KL}}}=\sum g_i \ln(g_i/r_i)$\\
\ Jeffreys & $d_{\mbox{\tiny{Jef}}}=\sum(g_i - r_i)\ln(g_i/r_i)$\\
\ K-divergence & $d_{\mbox{\tiny{Kdv}}}=\sum g_i \ln(2g_i/(g_i + r_i))$\\
\ Topsøe & $d_{\mbox{\tiny{Top}}}=\sum (g_i \ln\frac{2g_i}{g_i + r_i}+r_i \ln\frac{2r_i}{g_i + r_i})$\\
\  & $- (g_i + r_i)/2*ln((g_i + r_i)/2))$\\
\midrule
\textbf{Combination family} \\
\midrule
\ Taneja & $d_{\mbox{\tiny{Tan}}}=\sum (g_i + r_i)/2\ln(\frac{g_i + r_i}{2\sqrt{g_i r_i}})$\\
\ Kumar-Johnson & $d_{\mbox{\tiny{KJ}}}=\sum \frac{(g_i^2 - r_i^2)^2}{2(g_i r_i)^{3/2}}$\\
\ Average($L_1$-$L_\infty$) & $d_{\mbox{\tiny{avL}}}=\sum(|g_i - r_i| + \max_i|g_i - r_i|)/2$\\
\midrule
\textbf{Vicissitude family} \\
\midrule
\ Vicis-Wave\_Hedges & $d_{\mbox{\tiny{V}$_{wh}$}}=\sum |g_i - r_i|/\min(g_i,r_i)$ \\
\ Vicis-Symmetric $\chi^2_3$ & $d_{\mbox{\tiny{vs}${\chi^2_3}$}}=\sum (g_i - r_i)^2/\max(g_i,r_i)$ \\
\ Max-Symmetric $\chi^2$ & $d_{\mbox{\tiny{MaxS}}}=\max(\sum (g_i - r_i)^2/g_i,\sum (g_i - r_i)^2/r_i)$ \\
\ Min-Symmetric $\chi^2$ & $d_{\mbox{\tiny{MinS}}}=\min(\sum (g_i - r_i)^2/g_i,\sum (g_i - r_i)^2/r_i)$ \\
\bottomrule
\end{tabular}
}\\
\footnotesize{\textbf{Notes.} The functions used to compute distance or similarity measures require vectors X
and Y (in this case, the $g$- and $r$-band light curves) and return the corresponding similarities or distances per the various measures given above.}
\end{table*}

\subsection{Training with a heterogeneous labeled set}
\label{section:training_sets}
It is common practice to split the LS into two disjoint subsets: a training set -- usually the 70-75\% of the LS, used for model fitting -- and a validation set -- usually 30-25\% of the LS --, dedicated to the fine-tuning of the hyperparameters and to assessing the performance of the model during the learning process. 
Nevertheless, considering the heterogeneous nature of our sample of AGN that were selected on the basis of different properties, such a choice would lead to results strongly dependent on the type of AGN used in the training. \citet{decicco21} already addressed this issue, and thus, consistent with that work, we resorted to the leave-one-out cross-validation (LOOCV; \citealt{loocv}). This essentially consists in treating each source in the LS as a single-unit validation set, while the remaining sources in the LS constitute the training set and are therefore used to classify the excluded source. Once this is done for each of the sources in the LS, a prediction is available for each of them. Hence, the LS serves as both the training and as the validation set, as each single-unit is used in the validation phase when it is not included in the training set.

\section{Random forest-based tests using features in two bands}
\label{section:rf_tests}
As we mentioned in Sect. \ref{section:dataset}, what is new in this work with respect to \citet{decicco21} is the use of $g$-band observations in addition to and in combination with the $r$-band set of data. As a consequence, the first natural step is to compare the results obtained in \citet{decicco21} with the ones here obtained. We stress once again that \citet{decicco21} adopted for the validation the same approach used here, but in the present work we expanded the set of features compared to the one used in that work, consisting of the variability features computed from the $r$-band light curve, the morphology indicator, the five optical/NIR colors, and the MIR color, as detailed in Table \ref{tab:features}. We also point out that the $g$- and $r$-band light curves used in this work have the same number of points per source due to the data imputation procedure described in Sect. \ref{section:data_imputation}, the maximum number of points being 33.

In order to compare the two works and make further tests, here we analyzed the performance of a model trained via an RF algorithm. The code here used is based on the use of the Python \emph{scikit-learn} library. We took the imbalance in our classes into account by setting the parameter \texttt{class\_weight = balanced\_subsample} in the algorithm so that, for each bootstrap sample used to extract a subset of features to build a tree, the weights of each class were adjusted dynamically based on the class distribution in that bootstrap sample used to train that specific tree. 

In order to optimize the performance of our RF classifiers, we tested several possible combinations of the hyperparameters that typically affect the most the building process of the ensemble of decision trees and the way predictions are made. Specifically, we resorted to a grid search, which essentially requires as an input a grid of values for the various hyperparameters one aims at tuning, and then performs an exhaustive search over this grid to find their best combination via cross-validation and based on specific scoring metrics. Of course the possible combinations to test are infinite and the process is computationally expensive. Hence, one has to limit the number of possible values for each hyperparameter. Here we chose to tune the following hyperparameters:
\begin{itemize}
\item \textit{n\_estimators}: This defines the number of decision trees to be used to build a forest and, ideally, it should optimize the accuracy of the classification over a reasonable computational time; we tested the values 100, 300, 500;
\item \textit{min\_samples\_split}: This sets a minimum threshold for the number of objects required to split an internal node; we tested the values 2, 5, 10;
\item \textit{max\_depth}: This defines the depth of each tree, aiming at avoiding underfitting/overfitting; we tested the values 10, 20, 30, None, where the last one means that the tree ramification goes on until all leaves are pure (i.e., they only contain sources from one class) or until the stopping criterion defined by \textit{min\_samples\_split} is fulfilled; 
\item \textit{min\_samples\_leaf}: This sets the minimum number of objects required for a node not to be merged with its parent node, hence preventing an excessive growth of the tree; we tested the values 1, 2, 4;
\item \textit{max\_features}: This defines the number of features to consider for the splitting in each node; we tested the options \textit{sqrt} and \textit{log2}, respectively setting this number to the square root or the $\log_2$ of the total number of features.
\end{itemize}
Based on the chosen values, we tested 216 combinations for each classifier. We chose to evaluate the performance of our models via the balanced accuracy, which is essentially an average of recall for the positive and negative classes and thus means that we are taking into account that our LS is unbalanced. We report details about the obtained results in Appendix \ref{appendixA}.

Our classifiers were built making use of different sets of features for each test, but always including the morphology indicator and all the colors from Table \ref{tab:features}. 
\begin{itemize}
\item $D21$ test: This aimed at repeating what was done in \citet{decicco21}. Hence, the RF classifier made use of the same set of features there used; but, for the sake of consistency with the rest of the present work, we used the same LS -- made of 2543 instead of 2,414 sources -- which we introduced in Sect. \ref{section:icp}; this is the only case where we used the total number of points of the $r-$band light curves, the maximum being 54, and did not limit this number to 33 as in the rest of this work. Again, the choice of using the full set of points for each light curve was in order to be consistent with \citet{decicco21}. With this test we aimed at assessing how important the number of visits -- and hence of points in a light curve -- is in the AGN selection process.
\item $r$-band test: Here the RF classifier made use of the same set of features used in \citet{decicco21}, the LS consisting of the above-mentioned 2543 sources. The difference with the previous test in this list is that the light curves of the sources here used consist of up to 33 points, as we aimed at properly comparing the results from this classifier to the ones that we would obtain from $g$-band data; the light curves can include a maximum of four synthetic points, based on what we explained in Sect. \ref{section:data_imputation} and showed in Table \ref{tab:dataset}.
\item $g$-band test: The RF classifier made use of the same morphology indicator and colors used in the previous two tests, but the $r$-band features were replaced by the corresponding $g$-band features. We stress that, as in the previous test, the light curves of the sources used for this test consist of up to 33 points but, in this case, a maximum of 16 points can be synthetic. This means that, with this test, we were also trying to assess whether the nature (real or synthetic) of the light curve points affects the final classification.
\item $rg$ test: The RF classifier made use of all the features used in the previous two tests, i.e.: $r$-band features, $g$-band features, plus the morphology indicator and  colors introduced in Table~\ref{tab:features}. With this test we explored the option of using two bands (by using the synthetic points that we added with the imputation) for the selection process, instead of using only one.
\item $rg$ + bivariate feature test: The RF classifier made use of all the features selected for this work, that is, the ones used for the previous classifier plus the bivariate features reported in Table \ref{tab:bivariate}. This test was meant to assess the relevance of features combining simultaneous observations in the two bands used, in addition to features computed from single-band observations.
\item $(g-r)_{feat}$ test: The RF classifier made use of features defined as the difference between each $r$-band feature and the homologous $g$-band feature plus, as usual, the morphology indicator and the colors from Table \ref{tab:features}. This test investigated the variability of the ``colors'' of the various features initially defined for each band, thus identifying possible features that vary significantly from one band to another.
\item $(g-r)_{mag}$ test: The RF classifier made use of variability features computed from the light curves obtained as the magnitude difference between the $g$ and the $r$ band for each source plus, as usual, additional features, these being the morphology indicator and the colors from Table \ref{tab:features}. This test investigated the variability of the features obtained from ``color light curves''.
\end{itemize}

Table \ref{tab:confusion_matrices} reports some metrics that typically characterize the performance of a binary classifier, and that were derived from the confusion matrices obtained from the various classifiers tested in this work. We recall that, for a binary classifier, the confusion matrix consists of four frames, reporting the number of:
\begin{itemize}
\item[--] true positives (TPs), that is to say, known AGN correctly classified as AGN;
\item[--] true negatives (TNs), that is to say, known non-AGN correctly classified as non-AGN;
\item[--] false positives (FPs), that is to say, known non-AGN erroneously classified as AGN;
\item[--] false negatives (FNs), that is to say, known AGN erroneously classified as non-AGN.
\end{itemize}
Consistent with \citet{decicco21}, the metrics we used in this work are accuracy ($A$), precision ($P$, also known as purity), recall ($R$, also known as completeness), and $F1$, defined as follows:
\begin{equation*}
A = \frac{\mbox{TPs}+\mbox{TNs}}{\mbox{Tot. Sample}}\mbox{   , }
\end{equation*}
which tells how often the classification is correct for either class and is hence computed with respect to the whole sample of sources in the LS;
\begin{equation*}
P = \frac{\mbox{TPs}}{\mbox{TPs}+\mbox{FPs}}\mbox{   , }
\end{equation*}
which tells how often the classification as AGN is correct and therefore only refers to the sources classified as AGN;
\begin{equation*}
R = \frac{\mbox{TPs}}{\mbox{TPs}+\mbox{FNs}}\mbox{   , }
\end{equation*}
which tells how often known AGN are classified correctly and is hence computed with respect to all the known AGN in the LS;
\begin{equation*}
F1 = 2\times\frac{P\times R}{P+R}\mbox{   , }
\end{equation*}
which is the harmonic mean of $P$ and $R$ and therefore provides a different estimate of the accuracy, which also takes into account the sources for which the classification is wrong.

Specifically, the upper section of Table \ref{tab:confusion_matrices} refers to the various classifiers introduced above.
Since \citet{decicco21} also tested the use of various LSs, including different (sub)samples of AGN, we specify that the percentages reported in this work refer to the case where the full LS of 2543 sources is used. This is indeed the most interesting case if we consider that, in general, we do not know a priori what class of AGN we are dealing with, which depends on the specific properties of the survey and on the sample selection criteria.

\begin{table*}[tb]
\caption{\footnotesize{Confusion matrix values for the various classifiers tested.}} 
\label{tab:confusion_matrices}  
\renewcommand{\arraystretch}{1.5}
    \resizebox{.9\textwidth}{!}{
\begin{tabular}{l c c c c c c c}
\toprule
\ &                              TPR & TNR & accuracy & precision & unobscured AGN & obscured AGN & $F1$\\
\ &  (recall)                                &     &     &           & recall   & recall & \\
\hline
\ $r$, 54 visits   & $79.5\pm0.2$ & $98.76\pm0.06$ & $95.88\pm0.05$ & $91.8\pm0.3$ & $99.54\pm0.00$ & $48.4\pm0.6$ & $85.2\pm0.2$\\
\ $r$, 33 visits   & $79.6\pm0.9$ & $98.23\pm0.15$ & $95.45\pm0.22$ & $88.8\pm0.9$ & $98.6\pm0.4$ & $50.1\pm1.2$ & $83.9\pm0.8$\\
\ $g$,  33 visits & $80.0\pm0.5$ & $98.42\pm0.15$ & $95.67\pm0.13$ & $89.9\pm0.9$ & $98.2\pm0.3$ & $47.8\pm1.6$ & $84.7\pm0.4$\\
\ $rg$, 33 visits   & $76.8\pm0.5$ & $98.88\pm0.07$ & $95.59\pm0.08$ & $92.3\pm0.5$ & $98.2\pm0.0$ & $43.1\pm1.0$ & $83.9\pm0.3$\\
\ $rg + bivar.$, 33 visits & $73.3\pm0.7$ & $99.06\pm0.09$ & $95.21\pm0.14$ & $93.2\pm0.6$ & $97.9\pm0.2$ & $36.9\pm1.4$ & $82.0\pm0.6$\\
\ $(g-r)_{feat}$, 33 visits & $80.3\pm0.7$ & $98.34\pm0.09$ & $95.64\pm0.12$ & $89.4\pm0.5$ & $99.1\pm0.2$ & $48.7\pm1.6$ & $84.6\pm0.5$\\
\ $(g-r)_{mag}$, 33 visits  & $80.0\pm0.4$ & $97.84\pm0.08$ & $95.18\pm0.09$ & $86.7\pm0.4$ & $98.1\pm0.3$ & $50.0\pm1.5$ & $83.2\pm0.3$\\
\hline
\ $r$, 33 real, 0 synthetic    & $79.5\pm0.5$ & $98.30\pm0.06$ & $95.50\pm0.09$ & $89.2\pm0.4$ & $99.49\pm0.15$ & $48.4\pm1.4$ & $84.1\pm0.4$\\
\ $r$, 29 real, 4 synthetic   & $79.7\pm0.5$ & $98.16\pm0.08$ & $95.40\pm0.09$ & $88.4\pm0.5$ & $99.4\pm0.2$ & $48.5\pm1.0$ & $83.8\pm0.3$\\
\ $r$, 25 real, 8 synthetic  & $79.5\pm0.4$ & $98.07\pm0.12$ & $95.30\pm0.09$ & $87.9\pm0.6$ & $99.2\pm0.3$ & $48.2\pm1.5$ & $83.5\pm0.3$\\
\ $r$, 21 real, 12 synthetic  & $78.9\pm1.0$ & $98.29\pm0.15$ & $95.40\pm0.17$ & $89.0\pm0.9$ & $99.59\pm0.15$ & $45\pm2$ & $83.7\pm0.6$\\
\ $r$, 17 real, 16 synthetic  & $79.3\pm0.4$ & $98.23\pm0.11$ & $95.40\pm0.09$ & $88.7\pm0.6$ & $98.9\pm0.4$ & $46.3\pm1.2$ & $83.7\pm0.3$\\
\hline
\ $ks$, 25feat, 33 visits   & $82.2\pm0.6$ & $98.22\pm0.08$ & $95.82\pm0.11$ & $89.0\pm0.4$ & $98.7\pm0.2$ & $55\pm2$ & $85.5\pm0.4$\\          
\ $ks$, 9feat, 33 visits & $85.2\pm0.3$ & $97.75\pm0.10$ & $95.87\pm0.07$ & $86.9\pm0.5$ & $99.0\pm0.2$ & $66.7\pm1.0$ & $86.0\pm0.2$\\ 
\ $ks$, 8feat, 33 visits & $85.8\pm0.4$ & $97.55\pm0.11$ & $95.80\pm0.10$ & $86.0\pm0.5$ & $99.1\pm0.3$ & $68.1\pm1.2$ & $85.9\pm0.3$\\ 
\ $ks$, 7feat, 33 visits & $85.4\pm0.6$ & $97.54\pm0.14$ & $95.73\pm0.15$ & $85.9\pm0.7$ & $98.9\pm0.2$ & $67\pm2$ & $85.7\pm0.5$\\ 
\bottomrule
\end{tabular}
}\\
\\
\footnotesize{\textbf{Notes.} The table compares true positive ratio (TPR) and true negative ratio (TNR); accuracy, overall precision values, recall for unobscured and obscured AGN, and $F1$ values obtained in this work are also included, the overall value of the recall being the same as the TPR. All values are to be read as percent values. The percentage errors represent the standard deviation from the mean value derived from a set of ten simulations per classifier. In each simulation, the classifier builds a number of trees as detailed in \ref{appendixA} to determine the final classification for each source. What is new in this work is the introduction of additional features (see Sect. \ref{section:features}). The last four lines report the corresponding values obtained for four RF classifiers discussed in Sect. \ref{section:obscuredAGN}: they are built with the aim of optimizing the identification of obscured AGN in this work.}
\end{table*}

The analysis of the results obtained from this first series of tests led to a series of remarks, which we discuss in what follows:

\begin{itemize}
    \item The recall (quantified by the true positive ratio, TPR), that is, the largest fraction of correctly classified AGN, is generally consistent from test to test, except for the two classifiers combining $r$ and $g$ features; this also holds for the recall of obscured AGN, while we obtained a slightly higher value for the recall of unobscured AGN when we used 54 instead of 33 visits. The mild decrease in the TPR for the two classifiers $rg$ and $rg + bivar$ suggests that, if we aim at a higher recall, using only one band at a time is preferable, and also that the bivariate features we chose are not adding any relevant information to the tested classifiers.
    \item When combining the $r-$ and the $g-$band data ($rg$ and $rg + bivar$ classifiers), on the other hand, the precision we obtain is higher than in all the tests where only one band is used. This is due to a slightly higher TNR, and this result is consistent with several works from the literature showing that combining more bands usually returns less contaminated samples \citep[e.g.,][]{PSS21,Savic}, as well as with other yet-to-be-published results based on these same data where, combining three bands ($g$, $r$, and $i$), we obtain a purer sample of AGN candidates than the ones obtained from individual bands.
    \item Comparing the $g-$ and $r-$band tests, we find that the recall of either class of AGN is slightly lower for the $g$-band classifier, while all the other metrics are generally slightly higher for the g-band classifier. If we focus on the TPRs, there are 0.4\% more AGN correctly classified when using $g-$band features, which may suggest that, having a fixed number of 33 visits, the $g-$band dataset, where up to 16 points can be synthetic, is doing better than the $r$-band dataset, where only up to four points can be synthetic. In principle, this could be explained by the fact that we typically observe larger AGN variability in bluer bands \citep[e.g.,][]{Petrecca}. Nonetheless, we caution that we are comparing results in two different bands, one with mostly real visits and the other where about half the visits are synthetic. A more appropriate comparison would require making use of data from the same band to test the effect of the inclusion of synthetic visits, which we explore in the next section.
    \item The results from the test where color light curves are used are generally consistent with the others, except for the precision, which is the lowest in the whole upper section of the table.
    \item We stress that AGN recall, in general, is highly affected by the depth of the sample of sources under investigation. As a test, we computed the recall values for either class of AGN corresponding to different depths, obtained from the $gr$ classifier, and we found that, for unobscured AGN, we go from a 100\% recall value\footnote{We do not report uncertainties here when they are $<0.1\%$.} when the average $r-$band magnitude is $<20$ mag to a 97.0\% value for $<22$ mag, to the 98.2\% value reported in Table \ref{tab:confusion_matrices} for $<23.5$ mag; for obscured AGN we obtained 16.0\%, $(39.4\pm0.8)\%$, and $(43.1\pm1.0)\%$ for $<20$ mag, $<22$ mag, and $<23.5$ mag samples, respectively. In this last case, the large uncertainties are due to the small sizes of the available samples of obscured AGN (6, 46, and 104, respectively), which get larger with depth. This is a crucial point to keep in mind if we attempt a proper comparison of our results with other works from the literature, where the samples of AGN used are typically brighter than ours and thus mainly consist of unobscured AGN and therefore generally return recall values higher than ours \citep[e.g.,][]{PSS21}.
\end{itemize}

\subsection{Testing the impact of synthetic visits}
\label{section:imp_tests}
\begin{figure}
\centering
    {\includegraphics[width=9cm]{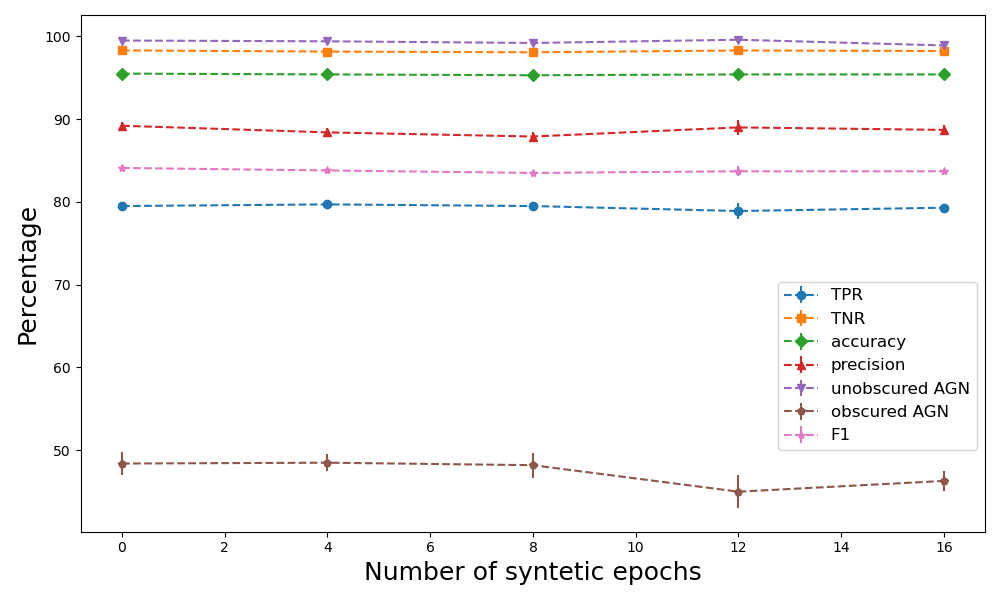}}
\caption{\footnotesize Comparison of the results obtained from the five tests where real visits were progressively replaced by synthetic visits for the various metrics used in this work. The total number of visits is always 33, and we replaced part of them, four by four, with synthetic visits, up to a maximum of 16. The only error bars large enough to be visible correspond to the recall for obscured AGN.}\label{fig:synth_vis_tests}
\end{figure}

As we mentioned in the previous section, when comparing the $r-$ and $g-$band classifiers, we are comparing results in two different bands where the datasets have different properties: one contains mostly real visits, while in the other about half of the visits are synthetic. Here we propose a more appropriate comparison making use of data from the same band, which allows us to test the impact of adding synthetic visits to a ``real'' dataset. With this in mind, we based our analysis on 33 visits, that is, the number of visits that we had been using for most of our tests so far. We therefore extracted a set of 33 visits from the original $r$-band dataset, which includes 54 visits. We selected them so as to cover the full baseline of 3.3 yr, and we required 16 of them to be replaceable with synthetic visits following the criterion described in Sect. \ref{section:data_imputation}, that is, if we exclude one of them, the time difference between the two adjacent visits must be $\leq 15$ days, so that we can replace that visit via linear interpolation. In this way, we could replicate with our $r-$band data a similar visit configuration to the one that we have for the $g$ band, with 17 real visits and 16 synthetic visits. Once identified this set of 33 real visits, we proceeded with the following tests, where we progressively replaced four, eight, 12, and 16 real visits with as many synthetic visits, building each time an RF classifier that made use of $r-$band features only:
\begin{itemize}
\item 33 real visits and no synthetic visits;
\item 29 real visits and four synthetic visits ($25\%$ of the maximum number of synthetic visits used in this work; this means that 12\% of the total number of visits are synthetic);
\item 25 real visits and eight synthetic visits ($50\%$ of the maximum number of synthetic visits used in this work; 24\% of the total number of visits are synthetic);
\item 21 real visits and 12 synthetic visits ($75\%$ of the maximum number of synthetic visits used in this work; 36\% of the total number of visits are synthetic);
\item 17 real visits and 16 synthetic visits (maximum number of synthetic visits used in this work; 48\% of the total number of visits are synthetic).
\end{itemize}
We report the results obtained from each of these tests in the middle section of Table \ref{tab:confusion_matrices}, as well as in Fig. \ref{fig:synth_vis_tests}, for a more immediate visualization. Essentially, we observe no noteworthy trend, which suggests that our imputation strategy is suitable to our purposes. There is a mild decrease in the recall of obscured AGN in the two bottom tests, as the fraction of synthetic visits increases, yet these values are consistent with the other ones in this section of the table within their uncertainties. For each feature we compared the distributions obtained when all the 33 visits are real to the corresponding distributions obtained when 16 out of 33 visits are synthetic. We resorted to the Kolmogorov-Smirnov (K-S) test to identify the pairs of distributions that changed the most; we found that, for seven of them, the distance $D$ returned from the test is $>0.78$ and the probability to obtain by chance a larger value is $<10^{-31}$. These features are, in descending order of D: \texttt{MaxSlope}, \texttt{$\eta^e$}, \texttt{IAR$_\phi$}, \texttt{GP\_DRW\_tau}, \texttt{LinearTrend}, \texttt{$R_{cs}$}, and \texttt{Period\_fit}. In Table \ref{tab:synth_test_feat_importance} we report for each of these seven features the position in the importance ranking obtained for the two cases that we are comparing (33 real visits versus 17 real and 16 synthetic visits), in order to show where these features place themselves and to assess whether the changes in their distributions affect the ranking. It is apparent that the features that are higher in the ranking when all 33 visits are real (\texttt{$\eta^e$}, \texttt{$R_{cs}$}, and \texttt{IAR$_\phi$}, which are in the top quartile) slide down to lower rankings when synthetic visits are introduced. This might be -- at least in part -- responsible for the mild drop that we observe for the recall of obscured AGN. Indeed, we anticipate that these features belong to the subset of features that we will select as the most suitable to identify obscured AGN; this is discussed in the next section, where we further investigate the issue of the identification of this class of AGN.

\begin{table}
\caption{\footnotesize{Position in the importance ranking for the seven features that were mostly affected by the replacement of 16 real visits with as many synthetic visits in order to test the impact of our data imputation strategy.}} 
\label{tab:synth_test_feat_importance}  
\centering
\footnotesize
\resizebox{.8\columnwidth}{!}{
\begin{tabular}{l c c}
\toprule
\ feature & ranking & ranking\\
\ & 33 real, 0 synthetic & 17 real, 16 synthetic \\
\ & visits & visits\\
\midrule
\ \texttt{MaxSlope}     & 26 & 24 \\
\ \texttt{$\eta^e$}     & 4  & 9  \\
\ \texttt{IAR$_\phi$}   & 9  & 19 \\
\ \texttt{GP\_DRW\_tau} & 8  & 10 \\
\ \texttt{LinearTrend}  & 18 & 15 \\
\ \texttt{$R_{cs}$}     & 6  & 17 \\
\ \texttt{Period\_fit}  & 21 & 31 \\
\bottomrule
\end{tabular}
}
\end{table}

\subsection{Selection of obscured AGN}
\label{section:obscuredAGN}
\begin{figure}[!hbtp]
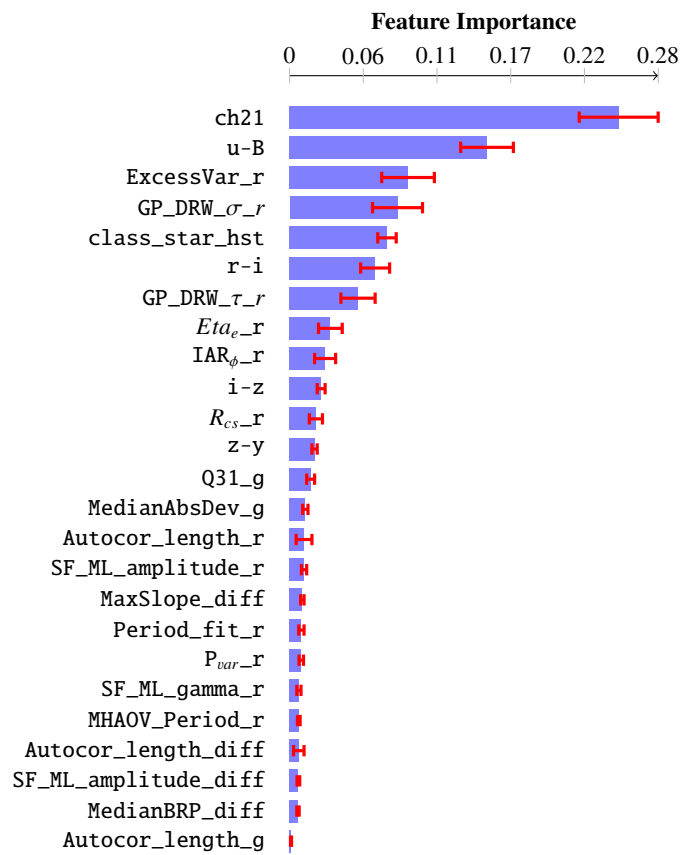

    \centering
\FeatureImportancePlot{
    \texttt{ch21} /0.25/0.03,
    \texttt{u-B}/0.15/0.02,
    \texttt{ExcessVar\_r}/0.09/0.02,
    \texttt{GP\_DRW\_$\sigma\_r$}/0.082/0.019,
    \texttt{class\_star\_hst}/0.074/0.007,
    \texttt{r-i}/0.065/0.011,
    \texttt{GP\_DRW\_$\tau\_r$}/0.052/0.013,
    \texttt{$Eta_e$\_r}/0.031/0.009,
    \texttt{IAR$_\phi$\_r}/0.027/0.008,
    \texttt{i-z}/0.024/0.003,
    \texttt{$R_{cs}$\_r}/0.020/0.005,
    \texttt{z-y}/0.019/0.002,
    \texttt{Q31\_g}/0.016/0.003,
    \texttt{MedianAbsDev\_g}/0.012/0.002,
    \texttt{Autocor\_length\_r}/0.011/0.006,
    \texttt{SF\_ML\_amplitude\_r}/0.011/0.002,
    \texttt{MaxSlope\_diff}/0.0097/0.0013,
    \texttt{Period\_fit\_r}/0.009/0.002,
    \texttt{P$_{var}$\_r}/0.0089/0.0017,
    \texttt{SF\_ML\_gamma\_r}/0.0071/0.0016,
    \texttt{MHAOV\_Period\_r}/0.0070/0.0010,
    \texttt{Autocor\_length\_diff}/0.007/0.004,
    \texttt{SF\_ML\_amplitude\_diff}/0.0067/0.0010,
     \texttt{MedianBRP\_diff}/0.0064/0.0009,
      \texttt{Autocor\_length\_g}/0.0011/0.0004
}
    \caption{Importance ranking for the top 25 features that, based on the results of the K-S test, allow a better separation between obscured AGN and inactive galaxies. }
    \label{fig:feat_imp_ks25}
\end{figure}

An interesting comparison concerns the recall of the samples of AGN retrieved using different classifiers: indeed, while it is well known that AGN selection based on optical variability is highly efficient in identifying unobscured AGN, it is also well known that it is generally not very effective in unearthing obscured AGN, as our series of VST-COSMOS works has widely proven \citep[][and references therein]{decicco15,decicco19,decicco21}; as a consequence, even a small improvement in the recall of the obscured AGN that we are able to retrieve via optical variability is relevant. 
The largest value here obtained so far for the recall of obscured AGN is $(50.1\pm1.2)\%$: though this is still much lower than the corresponding value obtained for unobscured AGN, it undoubtedly shows a significant improvement if compared to the initial 6\% value obtained in \citet{decicco15} with a five month baseline, or the 18\% value from \citet{decicco19}, where the baseline was the same as in this work, but the selection was based on a traditional approach, where we considered the r.m.s. deviation distribution and selected as variable AGN candidates all the sources with an r.m.s. deviation in excess of the 95\% percentile.\\
We know that, when trying to unearth obscured AGN, the main difficulty our classifiers have to face is separating them from inactive galaxies. With this in mind, we inspected the distributions of all the features used in this work, comparing via the K-S test the ones obtained for obscured AGN and the corresponding distributions for inactive galaxies, aiming at selecting those features that seem to better disentangle the two classes of sources. Based on the results of the K-S test, we selected the features where the distance between the two distributions in a pair is large and the corresponding probability to get by chance a larger distance is small. We identified a possible threshold for distances $D > 0.25$, which equals keeping 25 of the initial set of 162 features, and we built a classifier using only the selected features, which are reported in Table \ref{tab:sel_feat}. We note that no bivariate features are part of this selection; 12 out of 25 are $r-$band features, three are $g-$band features, four are obtained as differences between the two bands, five are colors, and one is the morphology indicator. Together with the classification, we obtained the feature importance ranking, shown in Fig. \ref{fig:feat_imp_ks25}. The importance for each feature was computed as the mean of the importance values that that feature has in each of the trees built by the classifier where that feature was used, where the total importance of a feature in a single tree is defined as the sum of the impurity reductions across all nodes where that feature is used. The importance of a feature is indeed strictly connected to the reduction of the impurity that results from using that feature to split the sample, thus generating a node. Error bars were obtained as the standard error associated with the mean value for each feature, and only the trees where the feature were used (i.e., were part of the bootstrapped sample of features) were included in the calculation.

Based on the obtained feature importance ranking, we proceeded as follows: we excluded from our new set of features the least important one, than tested a new classifier using the remaining 24 features. We repeated this procedure eliminating each time the last feature in the ranking and building a new classifier with all the other features. We stopped when we were left with seven features, as we noticed a trend in the results obtained test after test and, based on that, at some point we did not expect a further reduction in the number of features to lead to any improvements. We report the metrics obtained from the classifiers using 25 features plus the ones using nine to seven features, in descending order, in the bottom section of Table \ref{tab:confusion_matrices}. We omit the ones in between as the results are not particularly interesting.
What is apparent from this section of the table is that there is no classifier whose performance is consistently better than the others. Hence, the choice of the features to use for a model depends on the results we aim at. If our goal is obtaining a sample of obscured AGN that be as complete as possible, then we should base our selection on the eight features used in the classifier named \emph{ks8}. Indeed, in this case we managed to retrieve $(68.1 \pm 1.2)\%$ of known obscured AGN, a result that almost doubles the one obtained in \citet{decicco21}. Of course, as it usually happens, this higher recall comes at the expense of a higher contamination, reflected by the lower TNR and the corresponding lower precision. Specifically, we note that the TNR obtained from this classifier is 1.51\% lower than the highest value in the whole table, which corresponds to the \emph{rg~+ bivar} classifier. We note that the recall for unobscured AGN is consistent through the various \emph{ks} classifiers here discussed.

The feature importance ranking for the \emph{ks8} classifier is shown in Fig. \ref{fig:feat_imp_ks8}: we can see that the most important feature is, consistent with all the tests performed in this work as well as in \citet{decicco21}, the MIR color \texttt{ch21}; of the remaining features, three more are colors, one is the morphology indicator, and three are variability features. This final selection confirms once again the importance of combining light curves and color information to get improved results; see also Sect. 5.4 of \citet{decicco21}, which discusses the effects of using colors alone, without variability features, to select AGN via an RF classifier. We also note that the last $g$-band feature surviving the iterative reduction in the number of features used is \texttt{Q31\_g}, and it disappeared from the list when we were left with 12 features. From this point on, the only variability features used came from $r$-band light curves.

While in the bottom section of Table \ref{tab:confusion_matrices} the classifier with the highest TPR is \emph{ks8}, an overall look at the values obtained for TNR, precision, accuracy, and $F1$ shows that the \emph{ks9} classifier performs generally better than \emph{ks8}. Therefore, in principle, we could choose to keep one more feature -- namely, \texttt{IAR\_phi\_r} -- at the expense of selecting 1.4\% less obscured AGN. If we now look again at Table \ref{tab:confusion_matrices} as a whole, we can see that the reduction in the number of features used to build the various $ks$ classifiers does not imply a significant drop in the accuracy (-0.08\% if we compare the \emph{ks8} value to the highest accuracy value in the table, obtained from the \emph{r54} classifier), and it also returns a higher $F1$ while, as we mention above, the precision drops by 7.2\% if we compare the \emph{ks8} to the \emph{rg+bivar} classifier, or by 5.8\% if we compare the \emph{ks8} classifier to one-band only classifiers and select the one with the highest precision, namely the $r54$ classifier. Hence, once again, while this specific part of our work is focused on optimizing the selection of obscured AGN, in general, depending on the purpose and on how one plans to use the sample of sources classified as AGN, one might want to favor recall over contamination or vice versa.

\begin{figure}
\centering
    {\includegraphics[width=10cm]{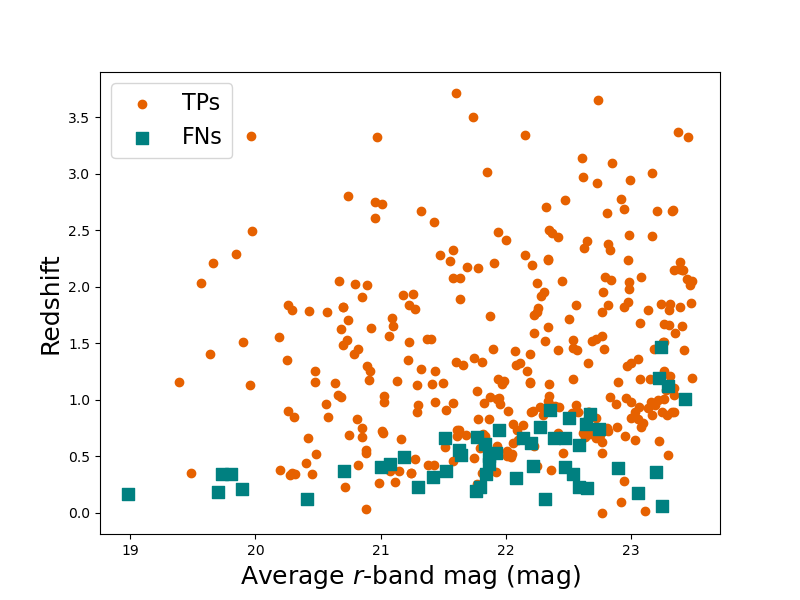}}
\caption{\footnotesize Redshift as a function of the average $r$-band magnitude for the AGN correctly classified (TPs, dark green dots) and the misclassified AGN (FNs, magenta squares) from the \emph{ks8} classifier. }\label{fig:z_vs_mag}
\end{figure}

Figure \ref{fig:z_vs_mag} shows redshift as a function of the average $r$-band magnitude for all the AGN in the LS, separating the ones correctly classified (TPs) from the misclassified ones (FNs), based on the results from the \emph{ks8} classifier. It is apparent that, while the two subsamples span quite uniformly the whole magnitude range covered by our dataset, the misclassified AGN are mainly lower-redshift sources and, based on what we showed in Fig. \ref{fig:hists} and also on the recall values reported in Table \ref{tab:confusion_matrices}, we know that these misclassified AGN are mostly obscured sources.

\begin{table}
\caption{\footnotesize{Features selected as the ones that better disentangle obscured AGN and inactive galaxies based on the results of the K-S test comparing, for each feature, the corresponding distributions for these two classes of sources.}} 
\label{tab:sel_feat}  
\footnotesize
\renewcommand{\arraystretch}{1.5}
\resizebox{.6\columnwidth}{!}{
\begin{tabular}{l c c}
\toprule
\ \textbf{feature} & \textbf{D} & \textbf{prob}\\
\midrule
\ \texttt{Autocor\_length\_r}    & 0.611 & $10^{-31}$\\
\ \texttt{SF\_ML\_amplitude\_r}  & 0.577 & $10^{-28}$\\
\ \texttt{SF\_ML\_gamma\_r}      & 0.547 & $10^{-25}$\\
\ \texttt{Autocor\_length\_g}    & 0.448 & $10^{-17}$\\
\ \texttt{i-z}                   & 0.376 & $10^{-12}$\\
\ \texttt{r-i}                   & 0.366 & $10^{-11}$\\
\ \texttt{MedianBRP\_diff}      & 0.353 & $10^{-10}$\\
\ \texttt{$R_{cs}$\_r}           & 0.334 & $10^{-9}$\\
\ \texttt{u-B}                   & 0.317 & $10^{-8}$\\ 
\ \texttt{z-y}                   & 0.315 & $10^{-8}$\\ 
\ \texttt{Autocor\_length\_diff} & 0.296 & $10^{-7}$\\
\ \texttt{ch21}                  & 0.295 & $10^{-7}$\\
\ \texttt{$Eta_e$\_r}            & 0.291 & $10^{-7}$\\ 
\ \texttt{class\_star\_hst}      & 0.287 & $10^{-7}$\\ 
\ \texttt{P$_{var}$\_r}          & 0.285 & $10^{-7}$\\  
\ \texttt{ExcessVar\_r}          & 0.283 & $10^{-7}$\\ 
\ \texttt{MHAOV\_Period\_r}       & 0.283 & $10^{-7}$\\ 
\ \texttt{GP\_DRW\_tau\_r}       & 0.281 & $10^{-7}$\\ 
\ \texttt{GP\_DRW\_sigma\_r}     & 0.275 & $10^{-6}$\\ 
\ \texttt{SF\_ML\_amplitude\_diff}  & 0.269 & $10^{-6}$\\
\ \texttt{Q31\_g}                & 0.267 & $10^{-6}$\\ 
\ \texttt{MedianAbsDev\_g}       & 0.262 & $10^{-6}$\\ 
\ \texttt{MaxSlope\_diff}         & 0.258 & $10^{-6}$\\ 
\ \texttt{Period\_fit\_r}         & 0.256 & $10^{-5}$\\ 
\ \texttt{IAR$_\phi$\_r}           & 0.251 & $10^{-5}$\\ 
\bottomrule
\end{tabular}
}\\
\\
\footnotesize{\textbf{Notes.} The features are listed in descending order of the distance D returned by the test, which represents the maximum distance between the two cumulative distributions compared; the table also reports the corresponding probability to obtain by chance a larger value for this distance.}
\end{table}

\begin{figure}[!hbtp]
    \centering
\FeatureImportancePlot{
    \texttt{ch21} /0.36/0.04,
    \texttt{u-B}/0.20/0.03,
    \texttt{class\_star\_hst}/0.103/0.009,
    \texttt{ExcessVar\_r}/0.10/0.03,
    \texttt{GP\_DRW\_$\sigma\_r$}/0.09/0.02,
    \texttt{r-i}/0.065/0.008,
    \texttt{GP\_DRW\_$\tau\_r$}/0.053/0.013,
    \texttt{i-z}/0.035/0.004
}
    \caption{Importance ranking for the eight features that were used to build the RF classifier identifying the highest fraction (i.e., returning the highest recall) of obscured AGN.}
    \label{fig:feat_imp_ks8}
\end{figure}

\section{Summary and conclusions}
\label{section:conclusions}
This study has evaluated the effectiveness of an RF classifier trained on various feature sets to identify AGN. The chosen features mostly characterize the optical variability of a source and were derived from light curves in two different bands, used individually, jointly, combined as bivariate features, and subtracted as ``color'' indicators for each feature or light curve. In particular, we focused on how to optimize the selection of obscured AGN, which are typically more challenging to detect through optical variability. Of course we do not expect any AGN photometric selection techniques that rely on optical variability to be able to return a completeness for obscured AGN that be anywhere near $100\%$, as we know that their optical emission should be at least in part hidden because of the presence of a dust torus or whatever structure might be responsible for the obscuration of the accretion disk. Nonetheless, in this work we show the way to extract sizable samples of obscured AGN, and we aim at testing this method on other datasets, keeping in mind that its efficiency will always depend on the amount of obscuration, intrinsic luminosity, and so on.

In order to draw broader conclusions from our analysis, and also considering the various findings from our series of studies using VST-COSMOS data \citep{decicco15,decicco19,decicco21}, we confirm the well-known fact that the observing cadence -- and consequently, the total number of visits used for selection -- is relevant, but we obtain comparable results from other classifiers. When defining the feature set obtained by single-band data alone, the best-performing algorithm among the ones tested is the only one that utilizes a larger (54) number of visits. In the various tests where the maximum number of visits is fixed to 33, the nature of the data points (real or synthetic) does not significantly impact the results, with slightly improved performance -- except for the recall -- when more synthetic points are added to the light curves ($g-$band test). However, tests examining the effect of synthetic points on real light curves suggest a possible decrease in the recall of obscured AGN as the fraction of synthetic to total number of points increases, while one important aim of this work is tailoring the selection method to increase the recall for obscured AGN.

Examining the three sections of Table \ref{tab:confusion_matrices} together reveals that the \emph{ks8} classifier returns the highest fraction of obscured AGN, this being $(68.1\pm1.2)\%$, at the expense of a precision that is of several percents lower than the values obtained from the classifiers using more features (upper section of Table \ref{tab:confusion_matrices}). This decrease in the precision originates from a larger contamination by false positives.
This test also confirms the crucial role of the only color based on MIR data, as well as of optical/NIR colors (three out of five are among the top eight features in the ranking), and also shows how $r-$band features dominate over $g-$band ones; a possible explanation for this could be, as mentioned, the larger presence of synthetic visits in the $g-$band dataset, but it can be also correlated with the fact that for obscured AGN bluer wavelengths, such as the $g$ band, are more absorbed with respect to the $r$ band.

Given the challenges we have faced in previous studies in the effort of retrieving larger fractions of obscured AGN, the achieved recall of $68.1\%$ indicates that building an accurate training sample and testing the optimal feature set is crucial to identify the obscured AGN population, and the feature set here identified could be valuable for further testing on different datasets, allowing us to assess their effectiveness in varying contexts, especially in view of wide-field surveys -- such as the already mentioned LSST -- which will provide us with much larger and richer source samples to investigate: indeed, with the LSST $ugrizy$ filters we will have densely sampled light curves in more bands than the ones we used in this work. In particular, the addition of the $izy$ filters will open up to a possible extension of our method to redder bands; if we focus once again on obscured AGN, typically enshrouded by dust and less affected by extinction compared to bluer bands, we expect their selection to benefit from the use of redder bands. This can therefore help detect variability that might be suppressed at the bluer wavelengths. In addition, analyzing their variability in the redder filters will allow us to better separate them from ``inactive'' galaxies, especially when combined with infrared data, which would allow detection of variability originating from dust reprocessing over longer scales. An important contribution in this wavelength regime is expected from the Euclid Mission (\citealt{Euclid2024}), with a plan for creating multiband catalogs of AGN and their host galaxies as a part of a set of Rubin-Euclid Derived Data Products (DDP; \citealp{Euclid-Rubin-DDP}). Another point is that, while our sample of obscured AGN does not extend to $z \gtrsim 1.5$, multiband variability can probe higher-redshift AGN as their variability signatures will be shifted into redder bands.\\
Our next goals while we wait for LSST data include the testing of our selection method optimized for the identification of obscured AGN on other datasets and, of course, once a sample of obscured AGN candidates is identified, we will need to validate it via other diagnostics and, when possible, via spectroscopic follow-up. In particular, we aim at testing our method over datasets of comparable or larger depth since, as also discussed in Sect. \ref{section:obscuredAGN}, a larger depth is necessary for the sample of obscured AGN to increase in size.

\begin{Acknowledgements}
DD acknowledges PON R\&I 2021, CUP E65F21002880003, and Fondi di Ricerca di Ateneo (FRA), linea C, progetto TORNADO. DD, MP, and VP acknowledge the financial contribution from PRIN-MIUR 2022 and from the Timedomes grant within the ``INAF 2023 Finanziamento della Ricerca Fondamentale''. SC acknoweldges the ASI-INAF TI agreement, 2018-23-HH.0 ``Attività scientifica per la missione Euclid - fase D'', and PRIN MUR 2022 (20224MNC5A), ``Life, death and after-death of massive stars'', funded by European Union – Next Generation EU.
\end{Acknowledgements}

\bibliographystyle{aa}
\bibliography{main}

\begin{thebibliography}{64}
\expandafter\ifx\csname natexlab\endcsname\relax\def\natexlab#1{#1}\fi

\bibitem[{{Allevato} {et~al.}(2013){Allevato}, {Paolillo}, {Papadakis}, \&
  {Pinto}}]{allevato}
{Allevato}, V., {Paolillo}, M., {Papadakis}, I., \& {Pinto}, C. 2013, \apj,
  771, 9

\bibitem[{{Amin} {et~al.}(2016){Amin}, {Anwar}, {Adnan}, {Nawaz}, {Howard},
  {Qadir}, {Hawalah}, \& {Hussain}}]{amin}
{Amin}, A., {Anwar}, S., {Adnan}, A., {et~al.} 2016, IEEE Access, 4, 7940

\bibitem[{{Antonucci}(1993)}]{Antonucci}
{Antonucci}, R. 1993, \araa, 31, 473

\bibitem[{{Botticella} {et~al.}(2017){Botticella}, {Cappellaro}, {Greggio},
  {Pignata}, {Della Valle}, {Grado}, {Limatola}, {Baruffolo}, {Benetti},
  {Bufano}, {Capaccioli}, {Cascone}, {Covone}, {De Cicco}, {Falocco},
  {Haeussler}, {Harutyunyan}, {Jarvis}, {Marchetti}, {Napolitano}, {Paolillo},
  {Pastorello}, {Radovich}, {Schipani}, {Tomasella}, {Turatto}, \&
  {Vaccari}}]{Botticella17}
{Botticella}, M.~T., {Cappellaro}, E., {Greggio}, L., {et~al.} 2017, \aap, 598,
  A50

\bibitem[{{Breiman}(2001)}]{Breiman2001}
{Breiman}, L. 2001, Machine Learning, 45, 5, cited By 34434

\bibitem[{{Brusa} {et~al.}(2010){Brusa}, {Civano}, {Comastri}, {Miyaji},
  {Salvato}, {Zamorani}, {Cappelluti}, {Fiore}, {Hasinger}, {Mainieri},
  {Merloni}, {Bongiorno}, {Capak}, {Elvis}, {Gilli}, {Hao}, {Jahnke},
  {Koekemoer}, {Ilbert}, {Le Floc'h}, {Lusso}, {Mignoli}, {Schinnerer},
  {Silverman}, {Treister}, {Trump}, {Vignali}, {Zamojski}, {Aldcroft},
  {Aussel}, {Bardelli}, {Bolzonella}, {Cappi}, {Caputi}, {Contini},
  {Finoguenov}, {Fruscione}, {Garilli}, {Impey}, {Iovino}, {Iwasawa},
  {Kampczyk}, {Kartaltepe}, {Kneib}, {Knobel}, {Kovac}, {Lamareille},
  {Leborgne}, {Le Brun}, {Le Fevre}, {Lilly}, {Maier}, {McCracken}, {Pello},
  {Peng}, {Perez-Montero}, {de Ravel}, {Sanders}, {Scodeggio}, {Scoville},
  {Tanaka}, {Taniguchi}, {Tasca}, {de la Torre}, {Tresse}, {Vergani}, \&
  {Zucca}}]{brusa}
{Brusa}, M., {Civano}, F., {Comastri}, A., {et~al.} 2010, \apj, 716, 348

\bibitem[{{Bruzual} \& Charlot(2003)}]{bc03}
{Bruzual}, G. \& Charlot, S. 2003, Monthly Notices of the Royal Astronomical
  Society, 344, 1000

\bibitem[{{Burbidge} {et~al.}(1963){Burbidge}, {Burbidge}, \& {Sandage}}]{BBS}
{Burbidge}, G.~R., {Burbidge}, E.~M., \& {Sandage}, A.~R. 1963, Reviews of
  Modern Physics, 35, 947

\bibitem[{{Capaccioli} \& {Schipani}(2011)}]{VST}
{Capaccioli}, M. \& {Schipani}, P. 2011, The Messenger, 146, 2

\bibitem[{{Cappellaro} {et~al.}(2015){Cappellaro}, {Botticella}, {Pignata},
  {Grado}, {Greggio}, {Limatola}, {Vaccari}, {Baruffolo}, {Benetti}, {Bufano},
  {Capaccioli}, {Cascone}, {Covone}, {De Cicco}, {Falocco}, {Della Valle},
  {Jarvis}, {Marchetti}, {Napolitano}, {Paolillo}, {Pastorello}, {Radovich},
  {Schipani}, {Spiro}, {Tomasella}, \& {Turatto}}]{Cappellaro15}
{Cappellaro}, E., {Botticella}, M.~T., {Pignata}, G., {et~al.} 2015, \aap, 584,
  A62

\bibitem[{{Cavuoti} {et~al.}(2024){Cavuoti}, {De Cicco}, {Doorenbos},
  {Brescia}, {Torbaniuk}, {Longo}, \& {Paolillo}}]{cavuoti24}
{Cavuoti}, S., {De Cicco}, D., {Doorenbos}, L., {et~al.} 2024, \aap, 687, A246

\bibitem[{Cha(2007)}]{Cha}
Cha, S.-H. 2007, Int. J. Math. Model. Meth. Appl. Sci., 1

\bibitem[{{De Cicco} {et~al.}(2021){De Cicco}, {Bauer}, {Paolillo}, {Cavuoti},
  {S{\'a}nchez-S{\'a}ez}, {Brandt}, {Pignata}, {Vaccari}, \&
  {Radovich}}]{decicco21}
{De Cicco}, D., {Bauer}, F.~E., {Paolillo}, M., {et~al.} 2021, \aap, 645, A103

\bibitem[{{De Cicco} {et~al.}(2022){De Cicco}, {Bauer}, {Paolillo},
  {S{\'a}nchez-S{\'a}ez}, {Brandt}, {Vagnetti}, {Pignata}, {Radovich}, \&
  {Vaccari}}]{decicco22}
{De Cicco}, D., {Bauer}, F.~E., {Paolillo}, M., {et~al.} 2022, \aap, 664, A117

\bibitem[{{De Cicco} {et~al.}(2015){De Cicco}, {Paolillo}, {Covone}, {Falocco},
  {Longo}, {Grado}, {Limatola}, {Botticella}, {Pignata}, {Cappellaro},
  {Vaccari}, {Trevese}, {Vagnetti}, {Salvato}, {Radovich}, {Brandt},
  {Capaccioli}, {Napolitano}, \& {Schipani}}]{decicco15}
{De Cicco}, D., {Paolillo}, M., {Covone}, G., {et~al.} 2015, \aap, 574, A112

\bibitem[{{De Cicco} {et~al.}(2019){De Cicco}, {Paolillo}, {Falocco},
  {Poulain}, {Brandt}, {Bauer}, {Vagnetti}, {Longo}, {Grado}, {Ragosta},
  {Botticella}, {Pignata}, {Vaccari}, {Radovich}, {Salvato}, {Covone},
  {Napolitano}, {Marchetti}, \& {Schipani}}]{decicco19}
{De Cicco}, D., {Paolillo}, M., {Falocco}, S., {et~al.} 2019, \aap, 627, A33

\bibitem[{{Donley} {et~al.}(2012){Donley}, {Koekemoer}, {Brusa}, {Capak},
  {Cardamone}, {Civano}, {Ilbert}, {Impey}, {Kartaltepe}, {Miyaji}, {Salvato},
  {Sanders}, {Trump}, \& {Zamorani}}]{donley}
{Donley}, J.~L., {Koekemoer}, A.~M., {Brusa}, M., {et~al.} 2012, \apj, 748, 142

\bibitem[{{Edelson} {et~al.}(1996){Edelson}, {Alexander}, {Crenshaw}, {Kaspi},
  {Malkan}, {Peterson}, {Warwick}, {Clavel}, {Filippenko}, {Horne}, {Korista},
  {Kriss}, {Krolik}, {Maoz}, {Nandra}, {O'Brien}, {Penton}, {Yaqoob},
  {Albrecht}, {Alloin}, {Ayres}, {Balonek}, {Barr}, {Barth}, {Bertram},
  {Bromage}, {Carini}, {Carone}, {Cheng}, {Chuvaev}, {Dietrich},
  {Dultzin-Hacyan}, {Gaskell}, {Glass}, {Goad}, {Hemar}, {Ho}, {Huchra},
  {Hutchings}, {Johnson}, {Kazanas}, {Kollatschny}, {Koratkar}, {Kovo}, {Laor},
  {MacAlpine}, {Magdziarz}, {Martin}, {Matheson}, {McCollum}, {Miller},
  {Morris}, {Oknyanskij}, {Penfold}, {Perez}, {Perola}, {Pike}, {Pogge},
  {Ptak}, {Qian}, {Recondo-Gonzalez}, {Reichert}, {Rodriguez-Espinoza},
  {Rodriguez-Pascual}, {Rokaki}, {Roland}, {Sadun}, {Salamanca}, {Santos-Lleo},
  {Shields}, {Shull}, {Smith}, {Smith}, {Snijders}, {Stirpe}, {Stoner}, {Sun},
  {Ulrich}, {van Groningen}, {Wagner}, {Wagner}, {Wanders}, {Welsh}, {Weymann},
  {Wilkes}, {Wu}, {Wurster}, {Xue}, {Zdziarski}, {Zheng}, \& {Zou}}]{Edelson}
{Edelson}, R.~A., {Alexander}, T., {Crenshaw}, D.~M., {et~al.} 1996, \apj, 470,
  364

\bibitem[{{Euclid Collaboration} {et~al.}(2024){Euclid Collaboration},
  {Mellier}, {Abdurro'uf}, {Acevedo Barroso}, {Ach{\'u}carro}, {Adamek},
  {Adam}, {Addison}, {Aghanim}, {Aguena}, {Ajani}, {Akrami}, {Al-Bahlawan},
  {Alavi}, {Albuquerque}, {Alestas}, {Alguero}, {Allaoui}, {Allen}, {Allevato},
  {Alonso-Tetilla}, {Altieri}, {Alvarez-Candal}, {Alvi}, {Amara}, {Amendola},
  {Amiaux}, {Andika}, {Andreon}, {Andrews}, {Angora}, {Angulo}, {Annibali},
  {Anselmi}, {Anselmi}, {Arcari}, {Archidiacono}, {Aric{\`o}}, {Arnaud},
  {Arnouts}, {Asgari}, {Asorey}, {Atayde}, {Atek}, {Atrio-Barandela}, {Aubert},
  {Aubourg}, {Auphan}, {Auricchio}, {Aussel}, {Aussel}, {Avelino},
  {Avgoustidis}, {Avila}, {Awan}, {Azzollini}, {Baccigalupi}, {Bachelet},
  {Bacon}, {Baes}, {Bagley}, {Bahr-Kalus}, {Balaguera-Antolinez}, {Balbinot},
  {Balcells}, {Baldi}, {Baldry}, {Balestra}, {Ballardini}, {Ballester},
  {Balogh}, {Ba{\~n}ados}, {Barbier}, {Bardelli}, {Baron}, {Barreiro},
  {Barrena}, {Barriere}, {Barros}, {Barthelemy}, {Bartolo}, {Basset},
  {Battaglia}, {Battisti}, {Baugh}, {Baumont}, {Bazzanini}, {Beaulieu},
  {Beckmann}, {Belikov}, {Bel}, {Bellagamba}, {Bella}, {Bellini}, {Benabed},
  {Bender}, {Benevento}, {Bennett}, {Benson}, {Bergamini}, {Bermejo-Climent},
  {Bernardeau}, {Bertacca}, {Berthe}, {Berthier}, {Bethermin}, {Beutler},
  {Bevillon}, {Bhargava}, {Bhatawdekar}, {Bianchi}, {Bisigello}, {Biviano},
  {Blake}, {Blanchard}, {Blazek}, {Blot}, {Bosco}, {Bodendorf}, {Boenke},
  {B{\"o}hringer}, {Boldrini}, {Bolzonella}, {Bonchi}, {Bonici}, {Bonino},
  {Bonino}, {Bonvin}, {Bon}, {Booth}, {Borgani}, {Borlaff}, {Borsato}, {Bosco},
  {Bose}, {Botticella}, {Boucaud}, {Bouche}, {Boucher}, {Boutigny}, {Bouvard},
  {Bouwens}, {Bouy}, {Bowler}, {Bozza}, {Bozzo}, {Branchini}, {Brando},
  {Brau-Nogue}, {Brekke}, {Bremer}, {Brescia}, {Breton}, {Brinchmann},
  {Brinckmann}, {Brockley-Blatt}, {Brodwin}, {Brouard}, {Brown}, {Bruton},
  {Bucko}, {Buddelmeijer}, {Buenadicha}, {Buitrago}, {Burger}, {Burigana},
  {Busillo}, {Busonero}, {Cabanac}, {Cabayol-Garcia}, {Cagliari}, {Caillat},
  {Caillat}, {Calabrese}, {Calabro}, {Calderone}, {Calura}, {Camacho Quevedo},
  {Camera}, {Campos}, {Canas-Herrera}, {Candini}, {Cantiello}, {Capobianco},
  {Cappellaro}, {Cappelluti}, {Cappi}, {Caputi}, {Cara}, {Carbone}, {Cardone},
  {Carella}, {Carlberg}, {Carle}, {Carminati}, {Caro}, {Carrasco}, {Carretero},
  {Carrilho}, \& {Carron Duque}}]{Euclid2024}
{Euclid Collaboration}, {Mellier}, Y., {Abdurro'uf}, {et~al.} 2024, \aap, in
  prep., arXiv:2405.13491

\bibitem[{{Eyheramendy} {et~al.}(2018){Eyheramendy}, {Elorrieta}, \&
  {Palma}}]{eyheramendy18}
{Eyheramendy}, S., {Elorrieta}, F., \& {Palma}, W. 2018, \mnras, 481, 4311

\bibitem[{{Falocco} {et~al.}(2015){Falocco}, {Paolillo}, {Covone}, {De Cicco},
  {Longo}, {Grado}, {Limatola}, {Vaccari}, {Botticella}, {Pignata},
  {Cappellaro}, {Trevese}, {Vagnetti}, {Salvato}, {Radovich}, {Hsu},
  {Capaccioli}, {Napolitano}, {Brandt}, {Baruffolo}, {Cascone}, \&
  {Schipani}}]{Falocco15}
{Falocco}, S., {Paolillo}, M., {Covone}, G., {et~al.} 2015, \aap, 579, A115

\bibitem[{{Fu} {et~al.}(2018){Fu}, {Liu}, {Radovich}, {Liu}, {Pan}, {Fan},
  {Covone}, {Vaccari}, {Amaro}, {Brescia}, {Capaccioli}, {De Cicco}, {Grado},
  {Limatola}, {Miller}, {Napolitano}, {Paolillo}, \& {Pignata}}]{Fu18}
{Fu}, L., {Liu}, D., {Radovich}, M., {et~al.} 2018, \mnras, 479, 3858

\bibitem[{{Graham} {et~al.}(2017){Graham}, {Djorgovski}, {Drake}, {Stern},
  {Mahabal}, {Glikman}, {Larson}, \& {Christensen}}]{graham17}
{Graham}, M.~J., {Djorgovski}, S.~G., {Drake}, A.~J., {et~al.} 2017, \mnras,
  470, 4112

\bibitem[{{Green} {et~al.}(2022){Green}, {Pulgarin-Duque}, {Anderson},
  {MacLeod}, {Eracleous}, {Ruan}, {Runnoe}, {Graham}, {Roulston}, {Schneider},
  {Ahlf}, {Bizyaev}, {Brownstein}, {del Casal}, {Dodd}, {Hoover}, {Matt},
  {Merloni}, {Pan}, {Ramirez}, {Ridder}, \& {Moseley}}]{Green22}
{Green}, P.~J., {Pulgarin-Duque}, L., {Anderson}, S.~F., {et~al.} 2022, \apj,
  933, 180

\bibitem[{{Guy} {et~al.}(2022){Guy}, {Cuillandre}, {Bachelet}, {Banerji},
  {Bauer}, {Collett}, {Conselice}, {Eggl}, {Ferguson}, {Fontana}, {Heymans},
  {Hook}, {Aubourg}, {Aussel}, {Bosch}, {Carry}, {Hoekstra}, {Kuijken},
  {Lanusse}, {Melchior}, {Mohr}, {Moresco}, {Nakajima}, {Paltani}, {Troxel},
  {Allevato}, {Amara}, {Andreon}, {Anguita}, {Bardelli}, {Bechtol}, {Birrer},
  {Bisigello}, {Bolzonella}, {Botticella}, {Bouy}, {Brinchmann}, {Brough},
  {Camera}, {Cantiello}, {Cappellaro}, {Carlin}, {Castander}, {Castellano},
  {Chari}, {Chisari}, {Collins}, {Courbin}, {Cuby}, {Cucciati}, {Daylan},
  {Diego}, {Duc}, {Fotopoulou}, {Fouchez}, {Gavazzi}, {Gruen}, {Hatfield},
  {Hildebrandt}, {Landt}, {Hunt}, {Ibata}, {Ilbert}, {Jasche}, {Joachimi},
  {Joseph}, {Knight}, {Kotak}, {Laigle}, {Lan{\c{c}}on}, {Larsen}, {Lavaux},
  {Leclercq}, {Leonard}, {von der Linden}, {Liu}, {Longo}, {Magliocchetti},
  {Maraston}, {Marshall}, {Mart{\'\i}n}, {Mattila}, {Maturi}, {McCracken},
  {Metcalf}, {Montes}, {Mortlock}, {Moscardini}, {Narayan}, {Paolillo},
  {Papaderos}, {Pello}, {Pozzetti}, {Radovich}, {Rejkuba}, {Rom{\'a}n},
  {S{\'a}nchez-Janssen}, {Sarpa}, {Sartoris}, {Schrabback}, {Sluse}, {Smartt},
  {Smith}, {Snodgrass}, {Talia}, {Tao}, {Toft}, {Tortora}, {Tutusaus}, {Usher},
  {van Velzen}, {Verma}, {Vernardos}, {Voggel}, {Wandelt}, {Watkins}, {Weller},
  {Wright}, {Yoachim}, {Yoon}, \& {Zucca}}]{Euclid-Rubin-DDP}
{Guy}, L.~P., {Cuillandre}, J.-C., {Bachelet}, E., {et~al.} 2022, in Zenodo id.
  5836022, Vol.~58, 5836022

\bibitem[{{Huijse} {et~al.}(2018){Huijse}, {Est{\'e}vez}, {F{\"o}rster},
  {Daniel}, {Connolly}, {Protopapas}, {Carrasco}, \&
  {Pr{\'\i}ncipe}}]{Huijse18}
{Huijse}, P., {Est{\'e}vez}, P.~A., {F{\"o}rster}, F., {et~al.} 2018, \apjs,
  236, 12

\bibitem[{{Ivezi{\'c}} {et~al.}(2019){Ivezi{\'c}}, {Kahn}, {Tyson}, {Abel},
  {Acosta}, {Allsman}, {Alonso}, {AlSayyad}, {Anderson}, {Andrew}, {Angel},
  {Angeli}, {Ansari}, {Antilogus}, {Araujo}, {Armstrong}, {Arndt}, {Astier},
  {Aubourg}, {Auza}, {Axelrod}, {Bard}, {Barr}, {Barrau}, {Bartlett}, {Bauer},
  {Bauman}, {Baumont}, {Bechtol}, {Bechtol}, {Becker}, {Becla}, {Beldica},
  {Bellavia}, {Bianco}, {Biswas}, {Blanc}, {Blazek}, {Bland ford}, {Bloom},
  {Bogart}, {Bond}, {Booth}, {Borgland}, {Borne}, {Bosch}, {Boutigny},
  {Brackett}, {Bradshaw}, {Brand t}, {Brown}, {Bullock}, {Burchat}, {Burke},
  {Cagnoli}, {Calabrese}, {Callahan}, {Callen}, {Carlin}, {Carlson}, {Chand
  rasekharan}, {Charles-Emerson}, {Chesley}, {Cheu}, {Chiang}, {Chiang},
  {Chirino}, {Chow}, {Ciardi}, {Claver}, {Cohen-Tanugi}, {Cockrum}, {Coles},
  {Connolly}, {Cook}, {Cooray}, {Covey}, {Cribbs}, {Cui}, {Cutri}, {Daly},
  {Daniel}, {Daruich}, {Daubard}, {Daues}, {Dawson}, {Delgado}, {Dellapenna},
  {de Peyster}, {de Val-Borro}, {Digel}, {Doherty}, {Dubois},
  {Dubois-Felsmann}, {Durech}, {Economou}, {Eifler}, {Eracleous}, {Emmons},
  {Fausti Neto}, {Ferguson}, {Figueroa}, {Fisher-Levine}, {Focke}, {Foss},
  {Frank}, {Freemon}, {Gangler}, {Gawiser}, {Geary}, {Gee}, {Geha}, {Gessner},
  {Gibson}, {Gilmore}, {Glanzman}, {Glick}, {Goldina}, {Goldstein}, {Goodenow},
  {Graham}, {Gressler}, {Gris}, {Guy}, {Guyonnet}, {Haller}, {Harris},
  {Hascall}, {Haupt}, {Hernand ez}, {Herrmann}, {Hileman}, {Hoblitt},
  {Hodgson}, {Hogan}, {Howard}, {Huang}, {Huffer}, {Ingraham}, {Innes},
  {Jacoby}, {Jain}, {Jammes}, {Jee}, {Jenness}, {Jernigan}, {Jevremovi{\'c}},
  {Johns}, {Johnson}, {Johnson}, {Jones}, {Juramy-Gilles}, {Juri{\'c}},
  {Kalirai}, {Kallivayalil}, {Kalmbach}, {Kantor}, {Karst}, {Kasliwal},
  {Kelly}, {Kessler}, {Kinnison}, {Kirkby}, {Knox}, {Kotov}, {Krabbendam},
  {Krughoff}, {Kub{\'a}nek}, {Kuczewski}, {Kulkarni}, {Ku}, {Kurita}, {Lage},
  {Lambert}, {Lange}, {Langton}, {Le Guillou}, {Levine}, {Liang}, {Lim},
  {Lintott}, {Long}, {Lopez}, {Lotz}, {Lupton}, {Lust}, {MacArthur}, {Mahabal},
  {Mand elbaum}, {Markiewicz}, {Marsh}, {Marshall}, {Marshall}, {May},
  {McKercher}, {McQueen}, {Meyers}, {Migliore}, {Miller}, {Mills}, {Miraval},
  {Moeyens}, {Moolekamp}, {Monet}, {Moniez}, {Monkewitz}, {Montgomery},
  {Morrison}, {Mueller}, {Muller}, {Mu{\~n}oz Arancibia}, {Neill}, {Newbry},
  {Nief}, {Nomerotski}, {Nordby}, {O'Connor}, {Oliver}, {Olivier}, {Olsen},
  {O'Mullane}, {Ortiz}, {Osier}, {Owen}, {Pain}, {Palecek}, {Parejko},
  {Parsons}, {Pease}, {Peterson}, {Peterson}, {Petravick}, {Libby Petrick},
  {Petry}, {Pierfederici}, {Pietrowicz}, {Pike}, {Pinto}, {Plante}, {Plate},
  {Plutchak}, {Price}, {Prouza}, {Radeka}, {Rajagopal}, {Rasmussen},
  {Regnault}, {Reil}, {Reiss}, {Reuter}, {Ridgway}, {Riot}, {Ritz}, {Robinson},
  {Roby}, {Roodman}, {Rosing}, {Roucelle}, {Rumore}, {Russo}, {Saha},
  {Sassolas}, {Schalk}, {Schellart}, {Schindler}, {Schmidt}, {Schneider},
  {Schneider}, {Schoening}, {Schumacher}, {Schwamb}, {Sebag}, {Selvy},
  {Sembroski}, {Seppala}, {Serio}, {Serrano}, {Shaw}, {Shipsey}, {Sick},
  {Silvestri}, {Slater}, {Smith}, {Smith}, {Sobhani}, {Soldahl},
  {Storrie-Lombardi}, {Stover}, {Strauss}, {Street}, {Stubbs}, {Sullivan},
  {Sweeney}, {Swinbank}, {Szalay}, {Takacs}, {Tether}, {Thaler}, {Thayer},
  {Thomas}, {Thornton}, {Thukral}, {Tice}, {Trilling}, {Turri}, {Van Berg},
  {Vanden Berk}, {Vetter}, {Virieux}, {Vucina}, {Wahl}, {Walkowicz}, {Walsh},
  {Walter}, {Wang}, {Wang}, {Warner}, {Wiecha}, {Willman}, {Winters},
  {Wittman}, {Wolff}, {Wood-Vasey}, {Wu}, {Xin}, {Yoachim}, \&
  {Zhan}}]{ivezich19}
{Ivezi{\'c}}, {\v{Z}}., {Kahn}, S.~M., {Tyson}, J.~A., {et~al.} 2019, \apj,
  873, 111

\bibitem[{Jeni {et~al.}(2013)Jeni, Cohn, \& De~la Torre}]{Jeni}
Jeni, L., Cohn, J., \& De~la Torre, F. 2013, in Proceedings - 2013 Humaine
  Association Conference on Affective Computing and Intelligent Interaction,
  ACII 2013, Vol. 2013

\bibitem[{{Kim} {et~al.}(2014){Kim}, {Protopapas}, {Bailer-Jones}, {Byun},
  {Chang}, {Marquette}, \& {Shin}}]{kim14}
{Kim}, D.-W., {Protopapas}, P., {Bailer-Jones}, C.~A.~L., {et~al.} 2014, \aap,
  566, A43

\bibitem[{{Kim} {et~al.}(2011){Kim}, {Protopapas}, {Byun}, {Alcock}, {Khardon},
  \& {Trichas}}]{kim11}
{Kim}, D.-W., {Protopapas}, P., {Byun}, Y.-I., {et~al.} 2011, \apj, 735, 68

\bibitem[{{Klesman} \& {Sarajedini}(2007)}]{Klesman&Sarajedini}
{Klesman}, A. \& {Sarajedini}, V. 2007, \apj, 665, 225

\bibitem[{{Koekemoer} {et~al.}(2007){Koekemoer}, {Aussel}, {Calzetti}, {Capak},
  {Giavalisco}, {Kneib}, {Leauthaud}, {Le F{\`e}vre}, {McCracken}, {Massey},
  {Mobasher}, {Rhodes}, {Scoville}, \& {Shopbell}}]{Koekemoer}
{Koekemoer}, A.~M., {Aussel}, H., {Calzetti}, D., {et~al.} 2007, \apjs, 172,
  196

\bibitem[{{Kormendy} \& {Richstone}(1995)}]{KR95}
{Kormendy}, J. \& {Richstone}, D. 1995, \araa, 33, 581

\bibitem[{{Laigle} {et~al.}(2016){Laigle}, {McCracken}, {Ilbert}, {Hsieh},
  {Davidzon}, {Capak}, {Hasinger}, {Silverman}, {Pichon}, {Coupon}, {Aussel},
  {Le Borgne}, {Caputi}, {Cassata}, {Chang}, {Civano}, {Dunlop}, {Fynbo},
  {Kartaltepe}, {Koekemoer}, {Le F{\`e}vre}, {Le Floc'h}, {Leauthaud}, {Lilly},
  {Lin}, {Marchesi}, {Milvang-Jensen}, {Salvato}, {Sanders}, {Scoville},
  {Smolcic}, {Stockmann}, {Taniguchi}, {Tasca}, {Toft}, {Vaccari}, \&
  {Zabl}}]{laigle}
{Laigle}, C., {McCracken}, H.~J., {Ilbert}, O., {et~al.} 2016, \apjs, 224, 24

\bibitem[{{LaMassa} {et~al.}(2015){LaMassa}, {Cales}, {Moran}, {Myers},
  {Richards}, {Eracleous}, {Heckman}, {Gallo}, \& {Urry}}]{lamassa15}
{LaMassa}, S.~M., {Cales}, S., {Moran}, E.~C., {et~al.} 2015, \apj, 800, 144

\bibitem[{{Liu} {et~al.}(2020){Liu}, {Deng}, {Fan}, {Fu}, {Covone}, {Vaccari},
  {Radovich}, {Capaccioli}, {De Cicco}, {Grado}, {Marchetti}, {Napolitano},
  {Paolillo}, {Pignata}, \& {Ragosta}}]{Liu20}
{Liu}, D., {Deng}, W., {Fan}, Z., {et~al.} 2020, \mnras, 493, 3825

\bibitem[{{Liu} {et~al.}(2018){Liu}, {Fu}, {Liu}, {Radovich}, {Wang}, {Pan},
  {Fan}, {Covone}, {Vaccari}, {Botticella}, {Capaccioli}, {De Cicco}, {Grado},
  {Miller}, {Napolitano}, {Paolillo}, \& {Pignata}}]{Liu18}
{Liu}, D., {Fu}, L., {Liu}, X., {et~al.} 2018, \mnras, 478, 2388

\bibitem[{{LSST Science Collaboration} {et~al.}(2009){LSST Science
  Collaboration}, {Abell}, {Allison}, {Anderson}, {Andrew}, {Angel}, {Armus},
  {Arnett}, {Asztalos}, {Axelrod}, \& et~al.}]{lsst}
{LSST Science Collaboration}, {Abell}, P.~A., {Allison}, J., {et~al.} 2009,
  ArXiv e-prints, arXiv:0912.0201

\bibitem[{{MacLeod} {et~al.}(2016){MacLeod}, {Ross}, {Lawrence}, {Goad},
  {Horne}, {Burgett}, {Chambers}, {Flewelling}, {Hodapp}, {Kaiser}, {Magnier},
  {Wainscoat}, \& {Waters}}]{macleod16}
{MacLeod}, C.~L., {Ross}, N.~P., {Lawrence}, A., {et~al.} 2016, \mnras, 457,
  389

\bibitem[{{Marchesi} {et~al.}(2016){Marchesi}, {Civano}, {Elvis}, {Salvato},
  {Brusa}, {Comastri}, {Gilli}, {Hasinger}, {Lanzuisi}, {Miyaji}, {Treister},
  {Urry}, {Vignali}, {Zamorani}, {Allevato}, {Cappelluti}, {Cardamone},
  {Finoguenov}, {Griffiths}, {Karim}, {Laigle}, {LaMassa}, {Jahnke}, {Ranalli},
  {Schawinski}, {Schinnerer}, {Silverman}, {Smolcic}, {Suh}, \&
  {Trakhtenbrot}}]{marchesi}
{Marchesi}, S., {Civano}, F., {Elvis}, M., {et~al.} 2016, \apj, 817, 34

\bibitem[{{McLaughlin} {et~al.}(1996){McLaughlin}, {Mattox}, {Cordes}, \&
  {Thompson}}]{mclaughlin}
{McLaughlin}, M.~A., {Mattox}, J.~R., {Cordes}, J.~M., \& {Thompson}, D.~J.
  1996, \apj, 473, 763

\bibitem[{{Nakos} {et~al.}(2009){Nakos}, {Willis}, {Andreon}, {Surdej},
  {Riaud}, {Hatziminaoglou}, {Garcet}, {Alloin}, {Baes}, {Galaz}, {Pierre},
  {Quintana}, {Page}, {Tedds}, {Ceballos}, {Corral}, {Ebrero}, {Krumpe}, \&
  {Mateos}}]{nakos}
{Nakos}, T., {Willis}, J.~P., {Andreon}, S., {et~al.} 2009, \aap, 494, 579

\bibitem[{{Nun} {et~al.}(2015){Nun}, {Protopapas}, {Sim}, {Zhu}, {Dave},
  {Castro}, \& {Pichara}}]{nun}
{Nun}, I., {Protopapas}, P., {Sim}, B., {et~al.} 2015, arXiv e-prints,
  arXiv:1506.00010

\bibitem[{{Petrecca} {et~al.}(2024){Petrecca}, {Papadakis}, {Paolillo}, {De
  Cicco}, \& {Bauer}}]{Petrecca}
{Petrecca}, V., {Papadakis}, I.~E., {Paolillo}, M., {De Cicco}, D., \& {Bauer},
  F.~E. 2024, \aap, 686, A286

\bibitem[{{Poulain} {et~al.}(2020){Poulain}, {Paolillo}, {De Cicco}, {Brandt},
  {Bauer}, {Falocco}, {Vagnetti}, {Grado}, {Ragosta}, {Botticella},
  {Cappellaro}, {Pignata}, {Vaccari}, {Schipani}, {Covone}, {Longo}, \&
  {Napolitano}}]{Poulain20}
{Poulain}, M., {Paolillo}, M., {De Cicco}, D., {et~al.} 2020, \aap, 634, A50

\bibitem[{{Rees}(1984)}]{Rees84}
{Rees}, M.~J. 1984, \araa, 22, 471

\bibitem[{{Richards} {et~al.}(2011){Richards}, {Starr}, {Butler}, {Bloom},
  {Brewer}, {Crellin-Quick}, {Higgins}, {Kennedy}, \& {Rischard}}]{richards11}
{Richards}, J.~W., {Starr}, D.~L., {Butler}, N.~R., {et~al.} 2011, \apj, 733,
  10

\bibitem[{{Salpeter}(1964)}]{Salpeter64}
{Salpeter}, E.~E. 1964, \apj, 140, 796

\bibitem[{Sammut \& Webb(2010)}]{loocv}
Sammut, C. \& Webb, G.~I., eds. 2010, Leave-One-Out Cross-Validation (Boston,
  MA: Springer US), 600--601

\bibitem[{{S{\'a}nchez-S{\'a}ez} {et~al.}(2023){S{\'a}nchez-S{\'a}ez},
  {Arredondo}, {Bayo}, {Ar{\'e}valo}, {Bauer}, {Cabrera-Vives}, {Catelan},
  {Coppi}, {Est{\'e}vez}, {F{\"o}rster}, {Hern{\'a}ndez-Garc{\'\i}a}, {Huijse},
  {Kurtev}, {Lira}, {Mu{\~n}oz Arancibia}, \& {Pignata}}]{PSS23}
{S{\'a}nchez-S{\'a}ez}, P., {Arredondo}, J., {Bayo}, A., {et~al.} 2023, \aap,
  675, A195

\bibitem[{{S{\'a}nchez-S{\'a}ez} {et~al.}(2019){S{\'a}nchez-S{\'a}ez}, {Lira},
  {Cartier}, {Miranda}, {Ho}, {Ar{\'e}valo}, {Bauer}, {Coppi}, \&
  {Yovaniniz}}]{sanchezsaez}
{S{\'a}nchez-S{\'a}ez}, P., {Lira}, P., {Cartier}, R., {et~al.} 2019, \apjs,
  242, 10

\bibitem[{{S{\'a}nchez-S{\'a}ez} {et~al.}(2021){S{\'a}nchez-S{\'a}ez}, {Reyes},
  {Valenzuela}, {F{\"o}rster}, {Eyheramendy}, {Elorrieta}, {Bauer},
  {Cabrera-Vives}, {Est{\'e}vez}, {Catelan}, {Pignata}, {Huijse}, {De Cicco},
  {Ar{\'e}valo}, {Carrasco-Davis}, {Abril}, {Kurtev}, {Borissova}, {Arredondo},
  {Castillo-Navarrete}, {Rodriguez}, {Ruz-Mieres}, {Moya},
  {Sabatini-Gacit{\'u}a}, {Sep{\'u}lveda-Cobo}, \&
  {Camacho-I{\~n}iguez}}]{PSS21}
{S{\'a}nchez-S{\'a}ez}, P., {Reyes}, I., {Valenzuela}, C., {et~al.} 2021, \aj,
  161, 141

\bibitem[{{Sarajedini} {et~al.}(2011){Sarajedini}, {Koo}, {Klesman}, {Laird},
  {Perez Gonzalez}, \& {Mozena}}]{sarajedini11}
{Sarajedini}, V.~L., {Koo}, D.~C., {Klesman}, A.~J., {et~al.} 2011, \apj, 731,
  97

\bibitem[{{Savi{\'c}} {et~al.}(2023){Savi{\'c}}, {Jankov}, {Yu}, {Petrecca},
  {Temple}, {Ni}, {Shirley}, {Kova{\v{c}}evi{\'c}}, {Nikoli{\'c}}, {Ili{\'c}},
  {Popovi{\'c}}, {Paolillo}, {Panda}, {{\'C}iprijanovi{\'c}}, \&
  {Richards}}]{Savic}
{Savi{\'c}}, {\DJ}.~V., {Jankov}, I., {Yu}, W., {et~al.} 2023, \apj, 953, 138

\bibitem[{{Schmidt} {et~al.}(2010){Schmidt}, {Marshall}, {Rix}, {Jester},
  {Hennawi}, \& {Dobler}}]{Schmidt}
{Schmidt}, K.~B., {Marshall}, P.~J., {Rix}, H.-W., {et~al.} 2010, \apj, 714,
  1194

\bibitem[{{Scoville} {et~al.}(2007{\natexlab{a}}){Scoville}, {Abraham},
  {Aussel}, {Barnes}, {Benson}, {Blain}, {Calzetti}, {Comastri}, {Capak},
  {Carilli}, {Carlstrom}, {Carollo}, {Colbert}, {Daddi}, {Ellis}, {Elvis},
  {Ewald}, {Fall}, {Franceschini}, {Giavalisco}, {Green}, {Griffiths}, {Guzzo},
  {Hasinger}, {Impey}, {Kneib}, {Koda}, {Koekemoer}, {Lefevre}, {Lilly}, {Liu},
  {McCracken}, {Massey}, {Mellier}, {Miyazaki}, {Mobasher}, {Mould}, {Norman},
  {Refregier}, {Renzini}, {Rhodes}, {Rich}, {Sanders}, {Schiminovich},
  {Schinnerer}, {Scodeggio}, {Sheth}, {Shopbell}, {Taniguchi}, {Tyson}, {Urry},
  {Van Waerbeke}, {Vettolani}, {White}, \& {Yan}}]{scoville}
{Scoville}, N., {Abraham}, R.~G., {Aussel}, H., {et~al.} 2007{\natexlab{a}},
  \apjs, 172, 38

\bibitem[{{Scoville} {et~al.}(2007{\natexlab{b}}){Scoville}, {Aussel}, {Brusa},
  {Capak}, {Carollo}, {Elvis}, {Giavalisco}, {Guzzo}, {Hasinger}, {Impey},
  {Kneib}, {LeFevre}, {Lilly}, {Mobasher}, {Renzini}, {Rich}, {Sanders},
  {Schinnerer}, {Schminovich}, {Shopbell}, {Taniguchi}, \&
  {Tyson}}]{scoville07b}
{Scoville}, N., {Aussel}, H., {Brusa}, M., {et~al.} 2007{\natexlab{b}}, \apjs,
  172, 1

\bibitem[{{Trevese} {et~al.}(2008){Trevese}, {Boutsia}, {Vagnetti},
  {Cappellaro}, \& {Puccetti}}]{Trevese}
{Trevese}, D., {Boutsia}, K., {Vagnetti}, F., {Cappellaro}, E., \& {Puccetti},
  S. 2008, \aap, 488, 73

\bibitem[{{Trevese} {et~al.}(1994){Trevese}, {Kron}, {Majewski}, {Bershady}, \&
  {Koo}}]{Trevese94}
{Trevese}, D., {Kron}, R.~G., {Majewski}, S.~R., {Bershady}, M.~A., \& {Koo},
  D.~C. 1994, \apj, 433, 494

\bibitem[{{Trevese} {et~al.}(1989){Trevese}, {Pittella}, {Kron}, {Koo}, \&
  {Bershady}}]{Trevese89}
{Trevese}, D., {Pittella}, G., {Kron}, R.~G., {Koo}, D.~C., \& {Bershady}, M.
  1989, \aj, 98, 108

\bibitem[{Tschopp \& Hernandez-Rivera(2017)}]{dist_meas}
Tschopp, M. \& Hernandez-Rivera, E. 2017, Quantifying Similarity and Distance
  Measures for Vector-Based Datasets: Histograms, Signals, and Probability
  Distribution Functions

\bibitem[{{Ulrich} {et~al.}(1993){Ulrich}, {Courvoisier}, \&
  {Wamsteker}}]{Ulrich93}
{Ulrich}, M.~H., {Courvoisier}, T.~J.~L., \& {Wamsteker}, W. 1993, \apj, 411,
  125

\bibitem[{{Urry} \& {Padovani}(1995)}]{Urry&Padovani}
{Urry}, C.~M. \& {Padovani}, P. 1995, \pasp, 107, 803

\bibitem[{{Vanden Berk} {et~al.}(2004){Vanden Berk}, {Wilhite}, {Kron},
  {Anderson}, {Brunner}, {Hall}, {Ivezi{\'c}}, {Richards}, {Schneider}, {York},
  {Brinkmann}, {Lamb}, {Nichol}, \& {Schlegel}}]{VandenBerk}
{Vanden Berk}, D.~E., {Wilhite}, B.~C., {Kron}, R.~G., {et~al.} 2004, \apj,
  601, 692

\end{thebibliography}

\newpage
\appendix
\section{Best hyperparameters obtained per classifier} \label{appendixA}
\begin{table}[!htbp]
\caption{\footnotesize{Set of best values obtained from a grid search-based optimization of the five hyperparameters that typically have the most influence in the performance of an RF classifier.}} 
\label{tab:hp}  
\renewcommand{\arraystretch}{1.5}
    \resizebox{\columnwidth}{!}{
\begin{tabular}{c c c c}
\toprule
\ RF classifier & n\_estimators & max\_depth & min\_samples\_split\\
\midrule
\ $r$, 54 visits   				 & 500 & 10 & 10\\
\ $r$, 33 visits				 & 100 & 10 & 10\\
\ $g$,  33 visits 				 & 300 & 20 & 2\\ 
\ $rg$, 33 visits   			 & 500 & 10 & 10\\ 
\ $rg + bivar.$, 33 visits & 500 & 10 & 10\\ 
\ $(g-r)_{feat}$, 33 visits 	 & 100 & 10 & 10\\ 
\ $(g-r)_{mag}$, 33 visits  	 & 500 & 10 & 10\\ 
\hline
\ $r$, 33 real, 0 synthetic      & 300 & 10  &  2\\ 
\ $r$, 29 real, 4 synthetic      & 300 & 10  & 10\\ 
\ $r$, 25 real, 8 synthetic      & 100 & 10  & 10\\ 
\ $r$, 21 real, 12 synthetic     & 100 & 10  &  2\\ 
\ $r$, 17 real, 16 synthetic     & 300 & 10  & 10\\ 
\hline
\ $rg$, 25feat, 33 visits        & 300 & 10  & 10\\ 
\ $rg$, 9feat, 33 visits        & 100 & 20  &  2\\ 
\ $rg$, 8feat, 33 visits        & 100 & 20  &  10\\ 
\ $rg$, 7feat, 33 visits        & 100 & 10  &  2\\ 
\bottomrule
\end{tabular}
}\\
\\
\footnotesize{\textbf{Notes.} The optimized hyperparameters are \emph{n\_estimators}, \emph{max\_depth}, \emph{min\_samples\_split}, \emph{min\_samples\_leaf}, and \emph{max\_features}; these were introduced in Sect. \ref{section:rf_tests}). The first column in the table lists the various classifiers tested in this work, whose performance metrics are reported in Table \ref{tab:confusion_matrices}. We do not include a column for the last two hyperparameters since our tests always returned the same best values $min\_samples\_leaf = 4$ and $max\_features =sqrt$, the only exception being the \emph{ks8} classifier, for which $max\_features =log2$.}
\end{table}

\end{document}